\documentclass[fleqn,usenatbib]{mnras}
\usepackage[british]{babel}
\usepackage{amsmath,amssymb,graphicx,calc,multicol}

\newcommand{\tmop}[1]{\ensuremath{\operatorname{#1}}}
\newcommand{\tmtextbf}[1]{{\bfseries{#1}}}
\newcommand{\tmfloatcontents}{}
\newlength{\tmfloatwidth}
\newcommand{\tmfloat}[5]{
  \renewcommand{\tmfloatcontents}{#4}
  \setlength{\tmfloatwidth}{\widthof{\tmfloatcontents}+1in}
  \ifthenelse{\equal{#2}{small}}
    {\setlength{\tmfloatwidth}{0.45\linewidth}}
    {\setlength{\tmfloatwidth}{\linewidth}}
  \begin{minipage}[#1]{\tmfloatwidth}
    \begin{center}
      \tmfloatcontents
      \captionof{#3}{#5}
    \end{center}
  \end{minipage}}

\title{Low radio frequency observations of seven nearby galaxies with GMRT}

\author[]{
Subhashis Roy $^1$ \thanks{email: roy@ncra.tifr.res.in} 
Souvik Manna $^1$ \\
$^{1}$National Center for Radio Astrophysics, TIFR, Pune University Campus, Ganeshkhind, Pune 411007, India\\}
\pubyear{2020}
\begin{document}
\label{firstpage}
\pagerange{\pageref{firstpage}--\pageref{lastpage}}
\maketitle

\begin{abstract}

We have observed seven nearby large angular sized galaxies at 0.33
GHz using GMRT with angular resolution of $\sim10''$ 
and sub-mJy sensitivity. Using archival higher frequency data at 1.4 or $\sim$6
GHz, we have then determined their spatially resolved non-thermal spectrum.
As a general trend, we find that the spectral indices are comparatively flat at the galaxy 
centres and gradually steepen with increasing galactocentric distances.
Using archival far infrared (FIR) MIPS 70 ${\mu} m$ data, we estimate the 
exponent of radio-FIR correlation. One of the galaxy (NGC 4826)
was found to have an exponent of the correlation of $\sim1.4$. Average exponent from
0.33 GHz data for the rest of the galaxies was 0.63$\pm$0.06 and is
significantly flatter than the exponent 0.78$\pm$0.04 obtained using 1.4
GHz data.
This indicates cosmic ray electron (CRe) propagation to
have reduced the correlation between FIR and 0.33 GHz radio. 
Assuming a model of simple isotropic diffusion of CRe, we find that the
scenario can explain the frequency dependent cosmic ray electron
propagation length scales for only two galaxies. Invoking streaming instability
could, however, explain the results for the majority of the remaining ones.

\end{abstract}


\section{Introduction:}

Emission from galaxies acts as a tracer of energetic processes. Different
physical processes cause emission to peak in different wavebands. Low radio
frequency emission mostly originates from non-thermal emission due to
acceleration of high energy cosmic-ray electrons (CRes) in the ambient 
galactic magnetic fields. Thus the diffuse radio non-thermal emission traces the journey
of the CRe from the production site to different parts of the galaxy. 
Cosmic ray production is linked to star formation rates
(SFRs) through supernovae, and stars act as the main source of energy in
galaxies. Therefore, studying low radio frequency emission at high 
angular resolution from different
parts of a galaxy could provide important information not only of SFRs, but
also on how the CRes diffuse from their place of origin and their interaction
with other components of the interstellar medium (ISM) and the strength of
magnetic fields.

There have been significant number of observational studies of radio emission
from nearby large galaxies and to characterise the propagation of CRes in
them. Generally the observed radio emission is a mixture of thermal
and non-thermal emission. 
Thermal emission in galaxies is mostly seen to originate from dense
ionised gas in star forming regions (typically in galactic arms), and the nonthermal
emission arises from acceleration of CRes originating in supernova shocks 
\citep[e.g.,][]{vanderKruit1971A&A,vanderKruit1977A&A,Hummel1981A&A,Beck&Graeve1982A&A,Condon1992ARA&A,Berkhuijsen&Beck2003A&A,Paladino2009A&A,Beck2020A&A}.
These observations also established that the non-thermal spectral 
index is of about $-$0.5 close to star forming regions, which is flatter
than the spectral index of $\sim -$0.75 in non star-forming regions
which is expected from the model of CRe acceleration
\citep[e.g.,][]{Bell1978MNRAS,Biermann1993A&A}. As the CRes propagate away from their
origin, they lose energy by several processes including synchrotron, inverse
Compton and ionisation. Among these, at low radio frequencies as presented
here, energy loss by synchrotron process dominates from most parts of typical
nearby star forming galaxies \citep{Basu2015MNRAS}.  
CRes can propagate
away through physical processes like diffusion, streaming instability and outflow (advection).
Radio spectra becomes steeper for regions away from the
major star forming regions (e.g., interarm) due to higher relative loss of
energy from high energy CRes, whose energy goes down faster with time
\citep{Basu2012MNRAS}. 
The radio emission is not only confined to the galactic disk, but several
sensitive studies have shown
many spirals to have significant halo emission reaching kpc distances above the
disk
\citep[e.g.,][]{Hummel1989A&AS,Hummel1990A&A,Hummel1991IAUS,Rossa&Dettmar2003A&A,Irwin2012AJ,Krause2018A&A}.

In star forming regions, both UV photons and CRes are generated. The UV photons in turn can heat up dust which then re-radiates in the far infra-red (FIR) band
\citep[e.g.,][]{Helou1985ApJ,Condon1992ARA&A}. Radio emission originates
from CRes in presence of magnetic fields which increases in dense star forming
regions. The above physical parameters are believed to give rise to
the radio-FIR correlation, which is one of the tightest correlation in
astrophysics that holds over five orders of magnitude in both radio and FIR
luminosities \citep[e.g.,][]{Condon1992ARA&A,Yun2001ApJ}. This correlation holds for
wide morphological class of galaxies
\citep{Wunderlich1987A&AS,Dressel1988ApJ,Price1992ApJ} including normal galaxies.  
\cite{Condon1992ARA&A} have studied this correlation in normal galaxies in
global scale using FIR band IRAS 60 $\mu$m and 100 $\mu$m FIR flux densities
\citep{Helou1988} and 1.4 GHz radio flux densities and found excellent
correlation and a dispersion of less than 0.2 dex.  \cite{Yun2001ApJ} have
found this correlation between monochromatic far-infrared luminosity (60
$\mu$m) and 1.4 GHz radio luminosity.
Radio-FIR correlation also holds good at kpc scales
\citep[e.g.,][]{Murgia2005A&A,Tabatabaei2007A&Aa,Xu1992A&A}. Linear correlation was found
between thermal radio emission and warm dust emission for LMC
\citep{Hughes2006MNRAS}. Non-thermal radio emission and cool dust emission were
found to be correlated in non-linear fashion for M31 \citep{Hoernes1998A&A}.
\cite{Dumas2011AJ} found the exponent of the radio-FIR correlation to vary among
centre, arm, interarm and outer regions for the galaxy. 
Several models exist to explain the correlation starting from global to parsec
scales. Assuming galaxies to be optically thick, calorimeter model
\citep{Voelk1989A&A,Lisenfeld1996A&A} can explain global correlation for many galaxies but fails
to explain the local scale correlation. Non-calorimeter model by
\cite{Niklas1997A&A} considered diffusion, radiative decay and finite escape
probability of cosmic rays and it can explain local scale correlation for
optically thick and thin regime of dust heated by UV photons. Radio-FIR correlation in smaller
scales in galaxies are governed by CRe propagation scale \citep{Tabatabaei2013}. More recently,
\citet{Heesen2019} found different exponent in radio-FIR correlation for steep and flat spectrum
regions in four nearby spirals. They also studied the CRe propagation lengths
at 0.14 and 1.4 GHz in these galaxies.

Though a large number of studies on nearby spirals have been made, most of
theses studies were performed at frequencies of 1.4 GHz and above, or had low
resolution at lower frequencies (e.g., $\sim 1'$ with WSRT at 0.33 GHz). To
find the common properties of these galaxies and to compare them with different
galaxy types, observing a complete set of galaxies is needed with high
resolution at lower frequencies. 
For example at metre wavelengths, 
due to longer lifetime, CRes can propagate to a
few kpcs from their origin. As a result the spatially resolved
radio-FIR correlation at resolution
below 1 kpc could be destroyed. This effect can be tested via 
spatially resolved radio-FIR correlation at each radio frequency.
For galaxies within $\sim 10$ Mpc, 1 kpc 
corresponds to an angular scale of $\gtrsim 30''$, which can be measured with
an interferometer with a resolution of $\sim 10''$. As CRes can propagate
through diffusion, streaming instability and through advection,
the above in conjunction with higher frequency observations could
provide important clues for propagation of CRes in nearby galaxies
\citep{Heesen2018MNRAS,Vollmer2020}.
There have been only a few studies of spatially resolved
radio-FIR correlation in nearby large galaxies utilising radio data below 1
GHz.
The study by \citet{Basu2012ApJ} found the radio-FIR exponent between 0.33 GHz and 70 ${\mu}m$ of
four large nearby spirals to be flatter as compared to using 1.4 GHz data. However, 
their study included only four large spirals, and to generalise the above trends,
enlarging the sample size is required.
To facilitate such a study, we have made a sample of 46
galaxies within 11 Mpc. Seven of these galaxies have been observed with the Giant 
Metrewave Radio Telescope (GMRT) at
0.33 GHz as a pilot project, the results of which are presented in this paper.
Sample selection is presented in Sect. 2, observational details and data
reduction are described Sect.~3. In Sect.~4, we present the results of our
observations including radio maps at 0.33 and 1.4 GHz, their spatially resolved
spectrum, properties of the individual galaxies, radio-FIR correlation 
and a simple model of cosmic ray propagation.
Discussions are presented in Sect. 5, and conclusions are drawn in Sect.~6.

\section{Sample selection}

Spitzer Local Volume Legacy (LVL) is a complete volume limited sample of 258
galaxies within 11 Mpc \citep{Dale2009ApJ}. For most of these galaxies, UV,
H$\alpha$, optical and multi-frequency IR observations have been made. To
carry out Giant Metrewave Radio Telescope (GMRT) observation we selected 46
out of 258 galaxies of LVL sample based on the following criteria:

(i) To have large number of resolution elements across the galaxies, we
selected galaxies with angular size larger than $6'$.
(ii) To minimise the zero spacing missing flux
problem of an interferometer like GMRT with shortest spacing of $\sim50-100$m, we used an upper cut-off of $17'$ 
($\le$20\% missing flux at the shortest spacing for Gaussian intensity distribution). 
(iii) Declination greater than $- 45^{\circ}$, so that they are visible from the
telescope site for $> 5$ hours a day, which allows for better $\tmop{uv}$-coverage through Earth rotation synthesis.

\section{Observations and data analysis:}

\subsection{GMRT observation}

We observed 7 out of 46 selected galaxies with GMRT as pilot observations at
0.33 GHz with a bandwidth of 32 MHz. A total of 512 frequency channels
were used across the observing band and typical on source observing time was $\sim 3 - 5$ hours.
Details of observations are given in Table-1. 
For making final images the data was analysed in various methods as 
mentioned below. For the galaxy NGC 2683, data editing,
calibration, self-calibration and imaging was performed by 
using the pipeline SPAM (Source Peeling and Atmospheric Modelling by
\citet{Intema2014ascl}). In this case, 3C286 was used as the primary calibrator. 
For all other
sources, initial data editing and calibration were performed by following
the standard approach in Astronomical Image Processing
System (Aips), where 3C286 was used as the primary calibrator. Phase
calibration was performed using observation of a P-band calibrator chosen from VLA
calibrator manual such that its angular distance is within 15 degrees from the
target source. It was observed periodically every $\sim$30 minutes. The
calibrated data for NGC 3627, 4096, 4449 and 4490 were self calibrated and
imaged using SPAM. Imaging of NGC 4826 and 5194 were done manually in Aips.
The initial images of these two sources were improved by phase only
self-calibration. Finally, while doing A\&P self calibration, overall gains
were normalised to unity  with solution interval of
10 minutes. 
To make the final images of the extended images along with the compact sources
with better fidelity than the standard Clean algorithm, we employed a
variant of multi-resolution Clean \citep{Wakker1988} using the task IMAGR in Aips. From the
self-calibrated UV data, we made (i) multi-facet images of compact sources with
high resolution using a short UV cutoff such that the Central Square baselines
($\lesssim 1$~km) of GMRT \citep{Swarup1991} are not used during imaging.  Then
the significant Clean components are removed from the UV data used in stage
(ii).  A low resolution
image of the field is made in stage (iii) from the data produced in (ii) with UV range
corresponding to Central Square baselines. During Clean in step (iii), extended
Gaussian components of FWHM about 0.5 times of resolution of the 
image are used. Since the resolution of the image produced in (iii) is pretty
low, a single facet for the whole field could be used. In stage (iv), the Clean
components produced in stage (iii) are subtracted from the original
self-calibrated UV data. Using the low resolution Clean component subtracted UV
data, another set of multi-facet images are made in stage (v) without any short UV
cutoff. Since most of the extended sources are removed through low resolution imaging in stage (iii),
these images do not show the typical instability exhibited by the Clean algorithm for extended sources.
To produce the final image, the Gaussian Clean components subtracted in stage
(iv) are added back to the image produced in stage (v). 
The technique described above uses a single facet for low resolution Clean with
cell-size of $\sim4$ smaller than the low resolution beam (stage iii). 
However, IMAGR in Aips uses the same cell size in low resolution Clean as is
done in high resolution. Use of a much larger cell-size in low resolution Clean
significantly speeds up the deconvolution process as compared to the one in IMAGR.
To measure rms noise in the images made, we used the Aips task IMEAN, which
fits a Gaussian noise profile to histogram made using the pixel value
distribution. When the image concerned is small and was from public archive, we
estimated rms noise in a couple of boxes (using TVSTAT) with sizes similar to
the source and made on opposite sides of the same. The quoted value is the average of the two.

\begin{figure}
    \includegraphics[width=\columnwidth]{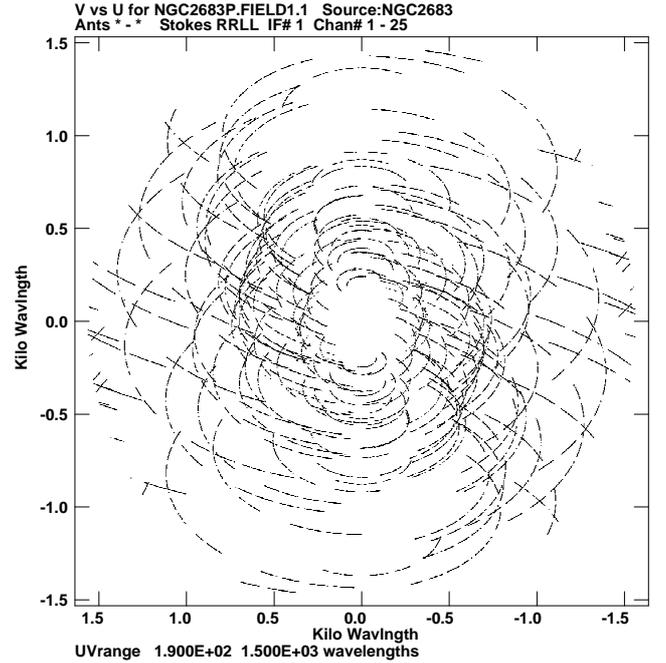}
 \caption{UV-coverage at 0.33 GHz within an $uv$-distance of 1.5 k$\lambda$ for the source NGC 2683.}
\end{figure}

\begin{table*}
\caption{Observation at 0.33 GHz with GMRT}
\centering
\begin{tabular}{||c c c c c c c||}
\hline
Source name & Observing date & On-source time & Bandwidth & No. of & Resolution & Rms \\
    
            &                & (hours)        & (MHz) & antennas working & ($'' \times ''$) &($\mu$Jy/beam)\\
\hline
    NGC 2683 & 11 Jan 2016 & 5.5 & 32 & 29 & $9.5 \times 7.2$ & 130 \\
    NGC 3627 & 14 Mar 2016 & 3.5 & 32 & 29 & $13.2 \times 8.5$& 450 \\
    NGC 4096 & 14 Mar 2016 & 3 & 32 & 29   & $10.8 \times 8.1$ & 60 \\
    NGC 4449 & 13 Mar 2016 & 4 & 32 & 29   & $10.6 \times 8.2$ & 80 \\
    NGC 4490 & 13 Mar 2016 & 4 & 32 & 29   & $10.3 \times 7.4$ & 70 \\
    NGC 4826 & 08 Jan 2016 & 4 & 32 & 28   & $ 9.2 \times 7.4$ & 170 \\
    NGC 5194 & 08 Jan 2016 & 4 & 32 & 28   & $ 9.5 \times 7.8$ & 100 \\
\hline
\end{tabular}
\end{table*}

\subsection{Higher frequency archival data}
To study propagation of CRes at higher frequencies (Sect.~4.3), to compare
radio-FIR correlation with 0.33 and 1.4 GHz radio data (Sect.~4.4) and to get
spectral indices for some of these galaxies we analysed VLA archival data for
all the seven galaxies at L-band ($\sim$1.4 GHz).
These data were
chosen such that the minimum bandwidth was of 50 MHz and observing time was more
than 20 minutes. Though the resolution of the GMRT images from UV data without any cutoffs are
high ($\sim 10 ''$, Table-1), to get reasonably good ($\gtrsim$ 5) signal to
ratio for extended emission of the galaxies (e.g., for making spectral index
images of extended emission), somewhat lower resolution was needed. 
Also, the propagation of CRes are studied through radio-FIR correlation (Sect.~ 4.3),
where 70 $\mu$m Spitzer FIR data has a resolution of 18$''$. Therefore, we
have chosen VLA L-band data such that we can get a resolution of $\sim 18''$.
This is typically obtained from data with C array. Therefore, we have made
images with either C array, C \& D array, or if C array data is not available
then from D and B array.
Data analysis were done using Aips following the standard technique of flagging bad data,
calibration using a secondary calibrator, self-calibration and imaging. 
Details of the data and images obtained are provided in Table-2
along with distances, position and inclination angles of the galaxies.
Rms noise
of NGC 4490 image made from VLA 1.4 GHz continuum data was significantly higher
than expected. Therefore, we used another data set (with the same project
code) of bandwidth 6 MHz towards the same object which had long observing time
of about 5 hours.  We analysed this line data set separately to calibrate and
extract the HI line free channels (36 out of 63 channels were used). A
continuum image was then made from the line free channels and the rms noise
achieved was very close to theoretical expectation.  The two continuum images
(one from the continuum data with 50 MHz bandwidth and another from the
6 MHz wide line data) were then averaged to get a better 1.4 GHz image of this
object with rms noise of 100 $\mu$Jy/beam. We have also used archival 4.8 GHz images 
of NGC 3627 \citep{Chyzy2000}, 4449 \citep{SOIDA2001} and  6 GHz ``{\it
uv}-tapered version'' \citep{Wiegert2015AJ} of NGC 2683 and 4096 to obtain and
study non-thermal spectral indices and CRe propagation in these galaxies.

\begin{table*}
\caption{Basic properties of the 0.33 and 1.4 GHz images}
 \centering 
  \begin{tabular} {||c c c c c c c c c||}
\hline
Source name & UV range &Resolution & GMRT Map rms & VLA Map rms & VLA Project & Distance & Position & Inclination \\
    
            &          &           & (0.33 GHz)    & (1.4 GHz) & (1.4 GHz)    &   of galaxies & angle & angle \\
    
           &  k$\lambda$ & ($'' \times ''$) & ($\mu$Jy/beam) & ($\mu$Jy/beam) &   & (Mpc) & (deg) & (deg)\\
\hline
  NGC 2683 & $0.19 - 15$ &$19 \times 13$ & 200 & 40 & AI23  & 7.7  &43 & 83\\
  NGC 3627 & $0.26 - 25$ &$16 \times 11$ & 800 & 370 & AS541 \& AP462 & 10.  &170 & 65\\
  NGC 4096 & $0.14 - 17$ &$14 \times 12$ & 100 & 25 & 16A-013 & 8.3  & 20 & 76\\
  NGC 4449 & $0.15 - 15$ &$26 \times 15$ & 300 & 180 & AB167 & 4.2  & 0 & 0\\
  NGC 4490 & $0.13 - 14$ &$19 \times 18$ & 230 & 100 & AA181 & 8.0  &126 & 60\\
  NGC 4826 & $0.22 - 20$ &$15 \times 14$ & 280 & 70 & AS541 & 7.5   & 120 & 60\\
  NGC 5194 & $0.15 - 10$ &$23 \times 18$ & 310 & 30 & AB505 \& AN57 & 8.0   & 10 & 20\\
\hline
  \end{tabular}
\end{table*}

\subsection{Separating thermal emission from radio maps}

Typical contribution of thermal emission is about 10\% at 1.4 GHz and $<$5\% at
0.33 GHz \citep{Basu2012ApJ}. However, this fraction could be $\sim$20-30\% at
1.4 GHz in high star forming regions.
Therefore, to study the properties of non-thermal emission from CRes,
separation of thermal emission is needed from the radio band maps.
H$\alpha$ is a direct tracer of ionised gas density. However, H$\alpha$ emission
is attenuated by dust and hence  H$\alpha$ maps needs to be extinction corrected.
Since the density of dust could be estimated from                       
their IR emission and correction can be made using this feature.
Dust absorption corrected H$\alpha$ emission has been used to
estimate the thermal emission in the past \citep{Tabatabaei2007A&A,Basu2012ApJ}. 
We use a similar technique as in \citet{Tabatabaei2007A&A}, where they have
estimated typical error in modelling the thermal emission could be $\sim20$\%.
While analysing images from different bands we have converted all images to a
common resolution and pixel size. 

We have used H$\alpha$ maps for six galaxies, NGC 2683, NGC 3627, NGC 4096, NGC 4449, NGC 4490, NGC 4826 
from 11HUGS \citep{Kennicutt2008ApJS}. 
For NGC 5194, H$\alpha$ map from Sings survey \citep{Kennicutt2003} was used.
H$\alpha$ maps have a resolution of 0.3$^{''}$ and flux density is recorded in
counts. To determine thermal emission, we 
have converted the unit of those maps from counts to erg/s/cm$^{2}$. 
\cite{Tabatabaei2007A&A} used 70 and 160 $\mu m$ FIR emission to correct for the 
dust extinction. We have used the same 
method using high resolution Herschel maps for the galaxies NGC 3627
\citep{Kennicutt2011}, NGC 4449 \citep{Madden2013,Karczewski2013} and NGC 4826
\citep{Kennicutt2011} at 70 and 160$\mu m$. We could make thermal maps of these
galaxies at the resolution of radio images using H$\alpha$ as a tracer. 
Herschel data at all the above bands are not available for rest of the galaxies.
We have used H$\alpha$ emission along with 24~$\mu m$ emission (see below) to
model the thermal emission from NGC 2683, NGC 4096 and NGC 4490.
We have used Spitzer MIPS 70 $\mu m$ (resolution 18$^{''}$) and 160 $\mu m$ (resolution 40$^{''}$) 
maps \citep{Dale2009ApJ} to model their thermal emission. However, the resolution of these maps
were limited to 40$^{''}$ as set by the 160 $\mu m$ MIPS maps. As described below, we used
24~$\mu$m as a tracer of thermal emission to model them at the resolution of radio images.

We estimated thermal emission from MIPS 24~$\mu m$ maps (resolution
6$^{''}$)  \citep{Rubin1968ApJ,Murphy2006ApJ} following the equation
$\frac{S_{\nu, th}}{ Jy.beam^{-1}} \sim 7.93 \times 10^{-3} T_{e4}^{0.45} \nu_{GHz}^{-0.1} [\frac{f_{24 \mu m}}{Jy.beam^{-1}}]$.
Here $T_{e4}$ is the electron temperature in the unit of $10^{4}$ K, $\nu_{GHz}$ is
the radio frequency in GHz and $f_{24\mu m}$ is the flux density at 24~$\mu$m.
As this emission is not a primary tracer of thermal emission, we have measured the galaxy
averaged ratio of thermal emission from H$\alpha$ as a tracer (see above) and from 24~$\mu$m
for NGC 2683, NGC 4096, NGC 4490 and NGC 5194. This ratio (correction
factor) is expected to depend on the galaxy type and is assumed to not vary
within a galaxy. Therefore, we have used the thermal maps from 24 $\mu m$ which
is multiplied by the correction factor.  This allowed us to achieve a resolution
comparable to radio maps. 

\subsection{Non-thermal spectral indices of the sample galaxies}

To make spectral index maps from available radio data, we aimed to
maximise signal to noise ratio.  With certain noises in images made at two
different radio frequencies, error
in spectrum is minimised if the two frequencies used are far apart.
Therefore, if sensitive images of the galaxies with the necessary short
spacing in $uv$ distance such that the Fourier transform of the Clean-components
from our P-band images show the missing flux density at the typical short
spacing ($\sim 500 \lambda$ corresponding to shortest spacing of 35m between
the antennas in D array) of VLA in C-band is within about 20\%,
we used thermal subtracted images of that band and P band ($\sim$0.33 GHz) to
make their non-thermal spectral index maps. 
Available C-band maps of NGC 2683 and 4096 \citep{Wiegert2015AJ}
satisfied the above criteria. C-band images of NGC 3627 \citep{Chyzy2000} and 4449
\citep{SOIDA2001} incorporated single dish data and did not suffer from 
missing flux densities at short ${uv}$ distances. We used the ``{\it uv}-tapered version''
\citep{Wiegert2015AJ} of NGC 2683 and 4096. 
We used the thermal subtracted maps of the above four galaxies to
make their non-thermal spectral index maps. For the rest of the galaxies NGC
4490, 4826 and 5194 we used thermal subtracted 1.4 GHz VLA images made for the
above. (Non-thermal spectral index images of NGC 2683, 3627, 4096 
and 4449 between 1.4 and 0.33 GHz and their thermal emission subtracted 1.4
GHz maps are provided in Supplementary part of this work.) To get the galaxy
structures represented adequately in both the frequency maps, we made
additional sets of images from the 0.33 GHz GMRT data matching the resolution
of the corresponding higher frequency images.
Typical UV-coverage obtained within upper cutoff of 1.5 $k\lambda$ which
provides the structure and intensity distribution of extended emission of sizes
of several arc-minutes have been shown for the source NGC 2683 at P-band with
GMRT. 

\section{Results:}

We first present the images and spatially resolved spectral index maps of the 
individual galaxies below and then study the cosmic ray propagation in these galaxies. 
Radio images made from GMRT 0.33 GHz data of the seven galaxies have been used to
generate contour plots (Table-1). As 24 $\mu m$ emission is a good tracer of thermal
emission, we have overlaid in gray scale the corresponding images of the
galaxies in IR at an wavelength of 24 $\mu m$ as observed by Spitzer. These
images have been presented below in Figs.~2 to 8. Unless mentioned separately,
the lowest contour plotted is 3 times the rms in all the plots presented in this work.
We have shown spatially resolved spectral index maps of the
galaxies in colour, where we have overlaid in contour the higher frequency
continuum images of the corresponding galaxies. For comparison, GMRT 0.33 GHz
maps of the galaxies at the same resolution are also presented.
These images have been presented below in Figs.~9 to 22. In spectral index
maps, pixels for which the signal to noise ratio was less than 5 were blanked. 
The absolute calibration error for GMRT 0.33 GHz data is typically about
$10$\%, while for the higher frequency data it is believed to be within
2$-$3\%. 

\subsection{Study of the individual galaxies}

(i) NGC 2683: It is a Sb type galaxy seen almost edge-on. This galaxy was
observed earlier at 5 GHz \citep{Sramek1975AJ}, and its integrated flux density was
measured to be 80$\pm 20$ mJy. Arecibo observations at 0.43 and 0.61 GHz
yielded its flux density as 0.34 Jy \citep{Lang&Terzian1969ApL}. More recent
observations with JVLA have shown its flux density to be 66.6$\pm$6.5 and
20.3$\pm$0.8 mJy at 19 and 5 cm respectively \citep{Wiegert2015AJ}. The
resultant galaxy integrated spectral index is $- 0.9$. The flux densities
measured earlier by single dishes could have been affected by confusion and
other sources in the field. Total flux density measured from our 0.33 GHz
observation is 188$\pm$10 mJy, and from the image made from archival VLA
C-array data at 1.4 GHz is 56$\pm$3 mJy. Galaxy integrated spectrum is $-
0.84 \pm 0.08$, which compares well with the spectrum determined from recent
observations at 4.8 and 1.4 GHz. Spectrum of the central core ($- 0.4 \pm
0.02$) and the galactic plane ($\simeq 0.4 - 0.5$) is significantly flatter
(Fig.~9) than the out of plane emission of the galaxy. \\
(ii) NGC 3627: It has a highly prominent bar with wide open arms (galaxy type
SAB) and is a member of the Leo group \citep{Dumas2007MNRAS}. This galaxy has
been observed before at 0.327 GHz with VLA with a resolution of 21.0$'' \times
16.6''$ \citep{Paladino2009A&A} and achieved an rms noise of 2 mJy. We have
achieved an rms noise of 0.8 mJy/beam, with a beam size of 16$'' \times
11''$ at 0.33 GHz. The measured flux density from our 0.33 GHz map is 1.7
$\pm 0.1$ Jy, which is almost double of $0.97 \pm 0.03$ Jy measured by them.
We note that there is a diffuse halo of size of about 6$'$ seen around this
galaxy, and the flux density for this object includes its contribution. It is
likely that due to a higher rms noise of the previous observation, the diffuse
extended emission was not seen, which could partly explain the discrepancy.
The L-band flux density from the archival VLA data is 0.33$\pm$0.01 Jy, which
is consistent with what is presented by \citet{Paladino2009A&A}. 
No significant diffuse emission could be seen
around this object in L-band map generated from B+D array VLA archival data.
They have tabulated its measured total flux densities from 57 MHz to 4.8 GHz
and plotted its integrated spectrum. Galaxy integrated spectrum from our
measurements at 1.4 and 0.33 GHz is $- 1.1 \pm 0.07$, which is way steeper
than what is estimated by them. It is likely caused by the contribution of the
halo seen at 0.33 GHz. The spectrum of the central compact emission ($- 0.48
\pm 0.03$) and the smaller sized source at the end of the southern bar ($-
0.59 \pm 0.02$) is flatter than the rest of the galaxy (see Fig.~11).\\
(iii) NGC 4096: It is a SABc type galaxy seen almost edge-on. Its flux density
measured by \citet{Condon1987ApJS} at 20 cm was 52.2 mJy. More recent
observations with JVLA have shown its flux density to be 57.1$\pm$1.1 and
16.3$\pm$0.3 mJy at 19 and 5 cm respectively \citep{Wiegert2015AJ}. From our
analysis, measured flux density for this galaxy is 174$\pm$10 and 56.2
$\pm$1.5 mJy from 0.33 GHz GMRT and 1.4 GHz JVLA archival data
respectively. Galaxy integrated spectrum is $- 0.78 \pm 0.06$ between 1.4
and 0.33 GHz. Emission from the central region has a spectral index of $- 0.64
\pm 0.02$, while that of the Northern compact source (Fig.~4) in the plane of
the galaxy is $- 0.5 \pm 0.02$ (Fig.~13). Their spectrum is flatter than rest
of the galaxy, indicating association of star forming regions with these
discrete emissions. \\
(iv) NGC 4449 is a dwarf irregular Magellanic type of galaxy. Three dense star
forming regions can be seen in its 24 $\mu m$ image (gray) overlaid on the
radio map. There is an extended halo seen in both IR and radio maps (Fig.~5).
The SNR 4449-1 is seen as a compact source embedded within the extended halo
of the galaxy towards North. Flux density of the galaxy in L-band was measured
to be 266 mJy \citep{Condon1987ApJS}. It has also been observed with GMRT at
0.61, 0.325 and 0.150 GHz \citep{Srivastava2014MNRAS}. Their measured flux
density for this galaxy at 0.33 GHz is 785$\pm$225 mJy, with an rms noise of
1.9 mJy. Our measured flux densities for this galaxy are 517$\pm$26 and
259$\pm$10 mJy from 0.33 GHz GMRT image and the image made from VLA 1.4 GHz
archival data respectively. The rms noise achieved from our 0.33 GHz data
after convolving to the archival L-band map is 0.3 mJy/beam (beam size of
26$'' \times 15''$), which is $\sim$5 lower than the previous
observation at 0.33 GHz. This yields galaxy integrated spectrum of $- 0.48 \pm
0.06$ between 1.4 and 0.33 GHz. Spectrum of the three dense star
forming regions in the south, north-west and north-east are $- 0.5 \pm 0.03$,
$- 0.25 \pm 0.02$ and $- 0.66 \pm 0.07$ respectively (Fig.~15).\\
(v) NGC 4490: This is a SBm type of galaxy, seen highly inclined.
This galaxy was observed in L-band using VLA \citep{Condon1987ApJS}, and
measured flux density was 774 mJy.  This galaxy has also been observed by GMRT
at 0.61 GHz by \citet{Nikiel-Wroczy2016MNRAS}
 and measured its total flux density to be 1426$\pm$116 mJy. Their
measured flux density of this interacting galaxy pair in L-band is 800$\pm$41
mJy. Our measured flux densities for this galaxy are 1990$\pm$100 and
850$\pm$20 mJy from our 0.33 GHz data and 1.4 GHz VLA archival data
respectively, which yields galaxy integrated spectrum of $- 0.59 \pm 0.07$.
Peak radio emission which appears to be the centre of the galaxy is offset from its peak 24
$\mu m$ emission (Fig.~6). Significant halo emission is seen in its 0.33 GHz
map, which has no counterpart in the 24 $\mu m$ image, indicating propagation
of CRe from the sites of production.
It is interacting with its smaller companion NGC 4485 seen towards North West.
Its spatially resolved spectral index map (Fig.~17) shows flat spectrum for the emission
along its plane. A separate flat spectrum region is also seen to extend from centre towards
East (seen along Dec 41$^{\circ}$39$'$ from centre), which could be an arm of
the galaxy or a region of triggered star formation due to merger of one of its
companion galaxy in the past \citep{Lawrence2020}\\
(vi) NGC 4826: This is a SAab type of galaxy with medium inclination. This
galaxy shows two counter-rotating gas-disks \citep{Braun1992Natur}. This galaxy was observed in 
L-band using VLA \citep{Condon1987ApJS}, and measured flux density to be 103 mJy.
Its flux density as measured by Sings survey is 110$\pm$10 mJy at 1.365 GHz
\citep{Braun2007A&A}. Our measured flux densities for this galaxy are 201
$\pm$11 and 99$\pm$3 mJy from 0.33 GHz GMRT image and the image made from VLA
1.4 GHz archival data. Galaxy integrated spectral index from our measurements
is $- 0.49 \pm .06$. The central part of the galaxy does show a flatter
spectrum of $- 0.4 \pm 0.07$ (Fig.~19). \\
(vii) NGC 5194: This is a Sbc type of grand design spiral galaxy seen nearly
face on. It is also interacting with its smaller companion NGC 5195 at its
North. This galaxy has been studied extensively in multiple wavebands and was
observed as a control sample due to previous observations of this object at
0.33 GHz with GMRT \citep{Mulcahy2014A&A,Mulcahy2016A&A}. Flux density of this galaxy as
measured in L-band is 1490 mJy \citep{Condon1987ApJS}. Our measured L-band flux
density from the archival C and D array VLA data is 1.2${\pm}$0.04 Jy. The
higher flux density of the earlier L-band image is likely due to its value
obtained from NRAO single dish map, which is not susceptible to missing
zero-spacing problem of an Interferometer. It could also result in part due to
a higher confusion noise of single dish measurements. We also compared its
flux density from our 0.33 GHz map with the one made earlier from GMRT data
using a factor of 2 smaller bandwidth \citep{Mulcahy2016A&A}. 
Its flux density as estimated by us from the previous GMRT image is
4.8 Jy, while the one imaged by us provides the flux density for the same
region as 3.7$\pm$0.2 Jy, which is about 30\% lower than what is measured from
the previous image. This discrepancy could not be due to a calibration error
as the flux densities of a few compact source in the field matched within a
few percents of each other in the previous map at the same frequency. The
difference is also unlikely from missing zero spacing, as our map was made
with short UV spacing above 150 $\lambda$ which provides the necessary flux
density for the size of NGC 5194 ($\sim 10'$). In fact, the same result was
obtained when we did imaging with UV spacing above 100 $\lambda$.  We believe
the difference arises from the way the 2 different images have been CLEAN-ed.

\begin{figure}
\includegraphics[width=\columnwidth]{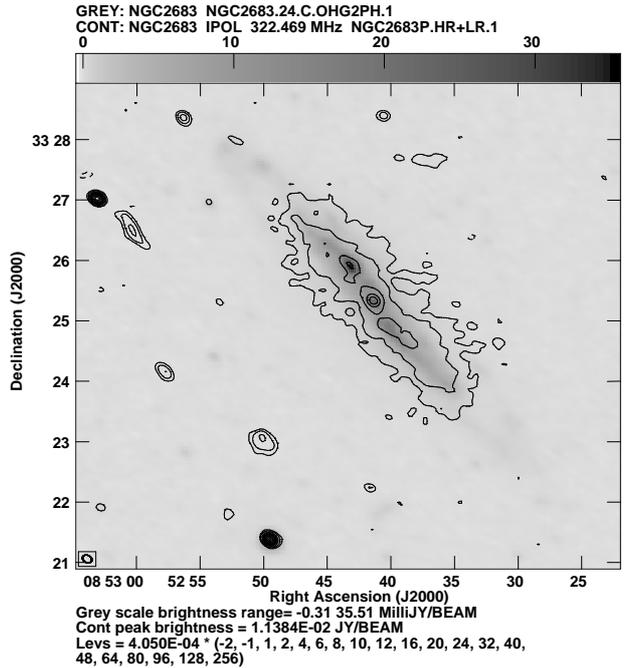}
\caption{NGC 2683 at 0.33 GHz (contour) and 24~$\mu$m (gray). Resolution 
is 9.5$^{''}\times 7.2^{''}$.}
\end{figure}

\begin{figure}
   \includegraphics[width=\columnwidth]{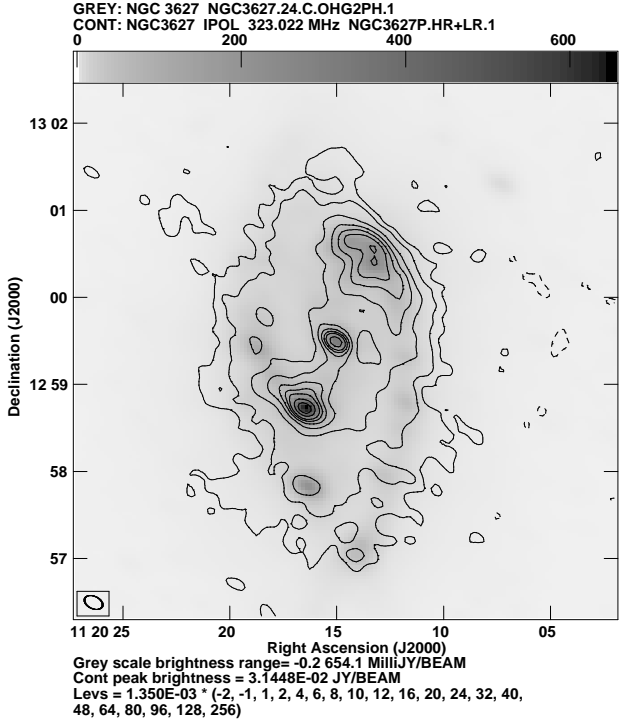}
\caption{NGC 3627 at 0.33 GHz (contour) and 24~$\mu$m (gray). Resolution 
is 13.2$^{''}\times 8.5^{''}$.}
\end{figure}

\begin{figure}
   \includegraphics[width=\columnwidth]{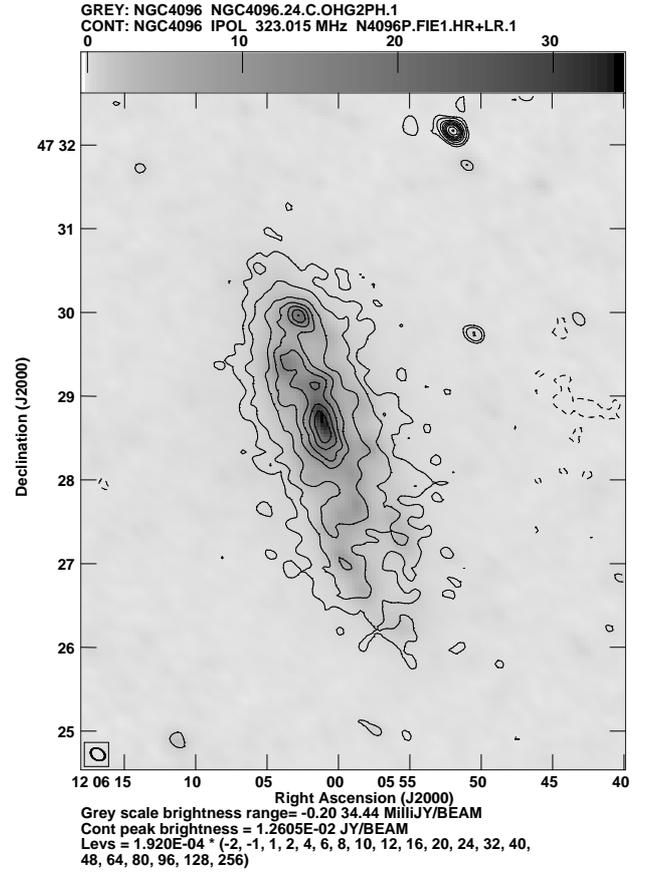}
\caption{NGC 4096 at 0.33 GHz (contour) and 24~$\mu$m (gray). Resolution 
is 10.8$^{''}\times 8.1^{''}$.}
\end{figure}

\begin{figure}
\includegraphics[width=\columnwidth]{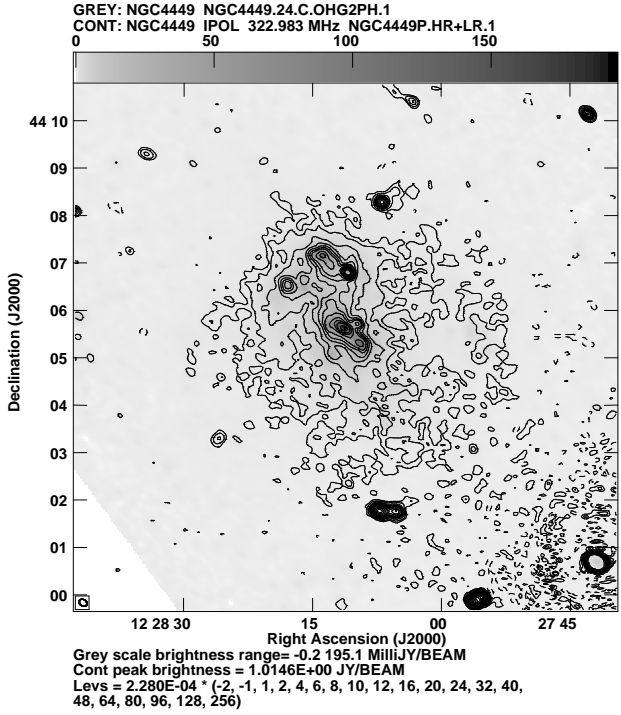}
\caption{NGC 4449 at 0.33 GHz (contour) and 24~$\mu$m (gray). Resolution 
is 10.6$^{''}\times 8.2^{''}$.}
\end{figure}

\begin{figure}
\includegraphics[width=\columnwidth]{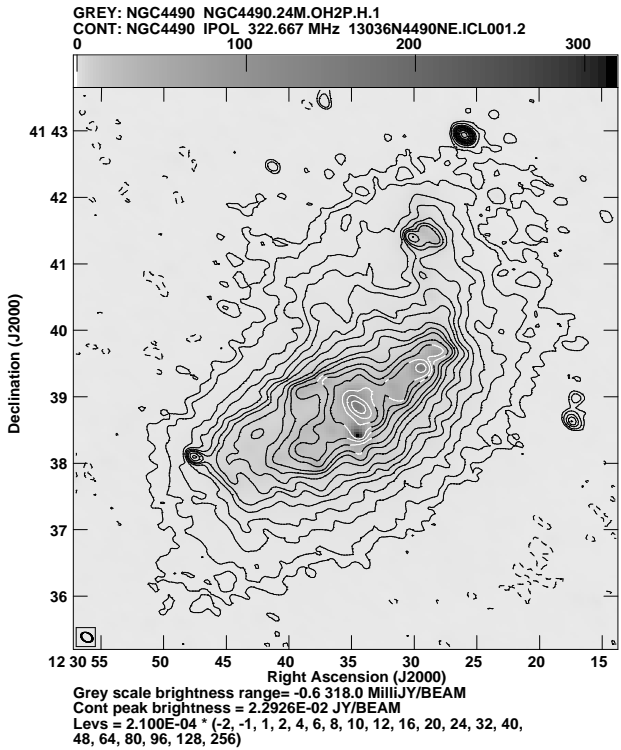}
\caption{NGC 4490 at 0.33 GHz (contour) and 24~$\mu$m (gray). Resolution 
is 10.3$^{''}\times 7.4^{''}$.}
\end{figure}

\begin{figure}
   \includegraphics[width=\columnwidth]{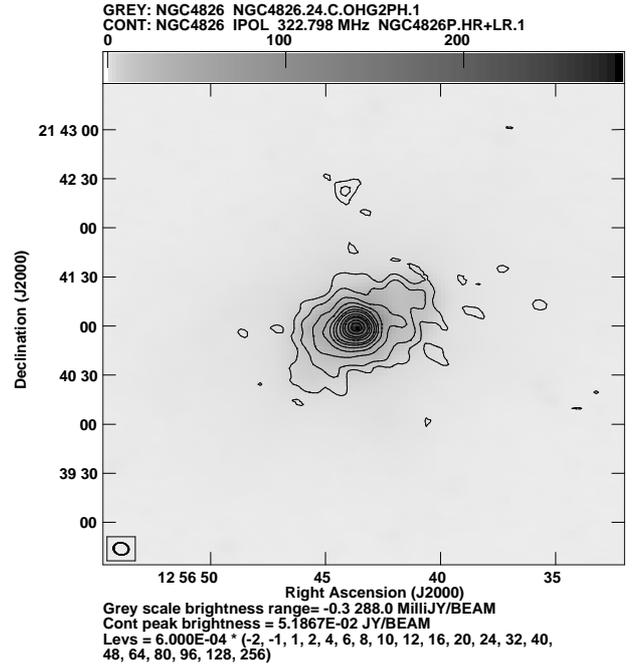}
\caption{NGC 4826 at 0.33 GHz (contour) and 24~$\mu$m (gray). Resolution 
is 9.2$^{''}\times 7.4^{''}$.}
\end{figure}

\begin{figure}
   \includegraphics[width=\columnwidth]{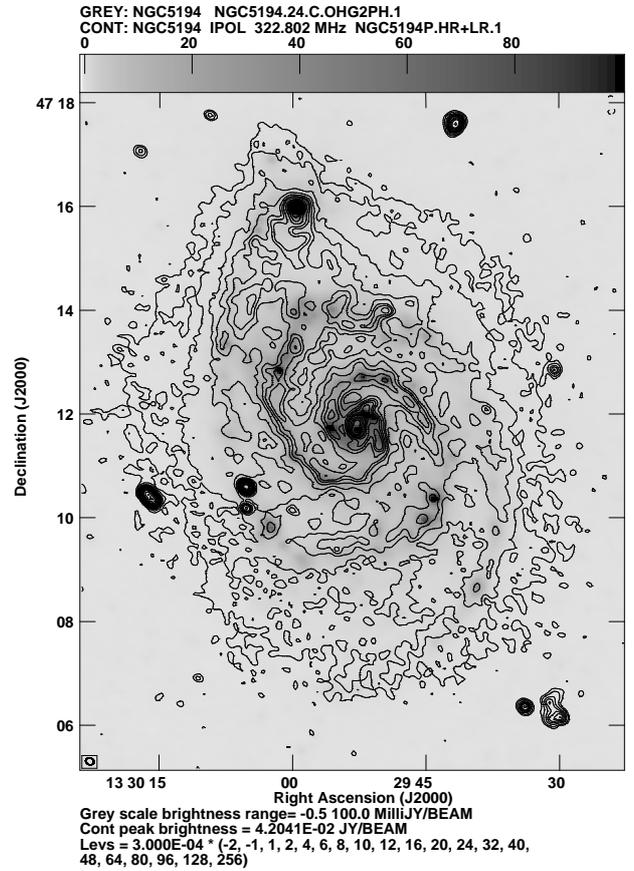}
\caption{NGC 5194 at 0.33 GHz (contour) and 24~$\mu$m (gray). Resolution 
is 9.5$^{''}\times 7.8^{''}$.}
\end{figure}

\begin{figure}
\includegraphics[width=\columnwidth]{N2683D_C.PBC.THS.Spix.col.ps}
\caption{Thermal emission subtracted 6 GHz map (contour) and non-thermal
spectral index of NGC 2683 (colour) between 0.33 and 6 GHz. Resolution is 15.9$^{''}\times 14.8^{''}$.}
\end{figure}.

\begin{figure}
\includegraphics[width=\columnwidth]{NGC2683P.ohg.6ghz.ths.ps}
\caption{Thermal emission subtracted 0.33 GHz map (in contour \& Gray scale) of NGC 2683.
Resolution is 15.9$^{''}\times 14.8^{''}$.}
\end{figure}.

\begin{figure}
   \includegraphics[width=\columnwidth]{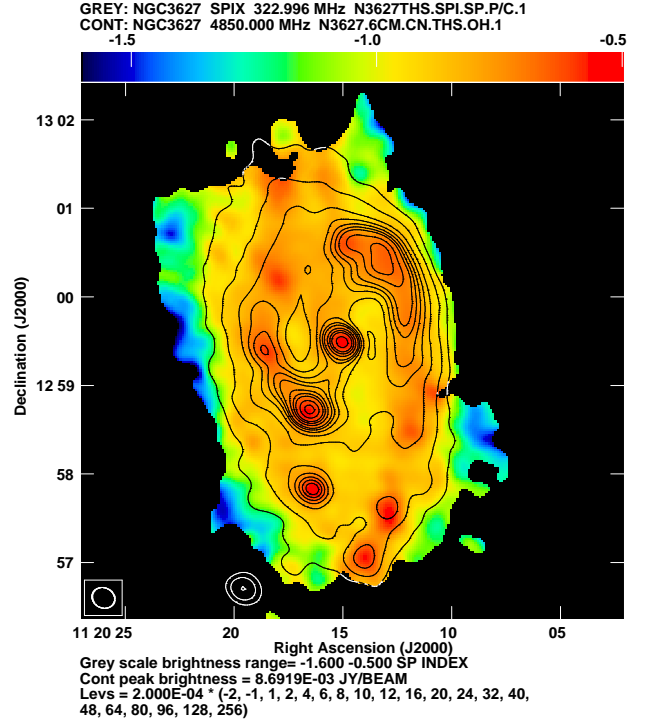}
\caption{Thermal emission subtracted 5 GHz map (contour) and non-thermal
spectral index of NGC 3627 (colour) between 0.33 and 5 GHz. Resolution is 15.7$^{''}\times 13.6^{''}$.}
\end{figure}

\begin{figure}
   \includegraphics[width=\columnwidth]{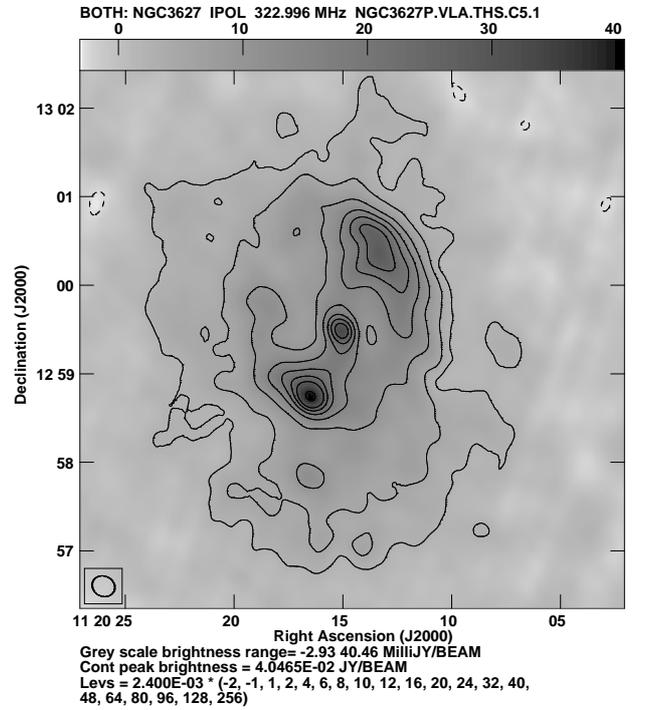}
\caption{Thermal emission subtracted 0.33 GHz map (in contour \& Gray scale) of NGC 3627.
Resolution is 15.7$^{''}\times 13.6^{''}$.}
\end{figure}.

\begin{figure}
\includegraphics[width=\columnwidth]{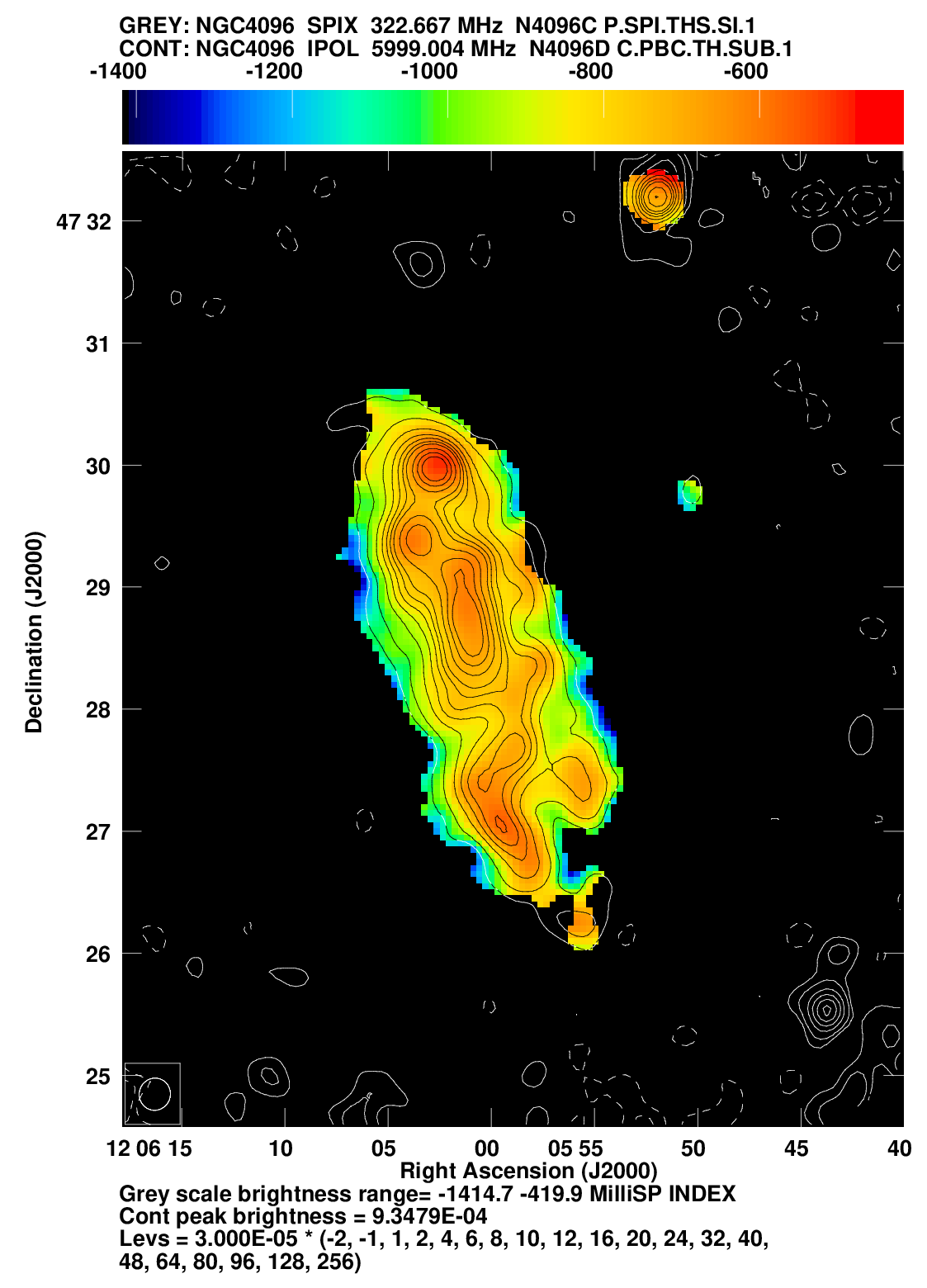}
\caption{Thermal emission subtracted 6 GHz map (contour) and non-thermal
spectral index of NGC 4096 (colour) between 0.33 and 6 GHz. Resolution is 15.9$^{''}\times 15.3^{''}$.}
\end{figure}

\begin{figure}
   \includegraphics[width=\columnwidth]{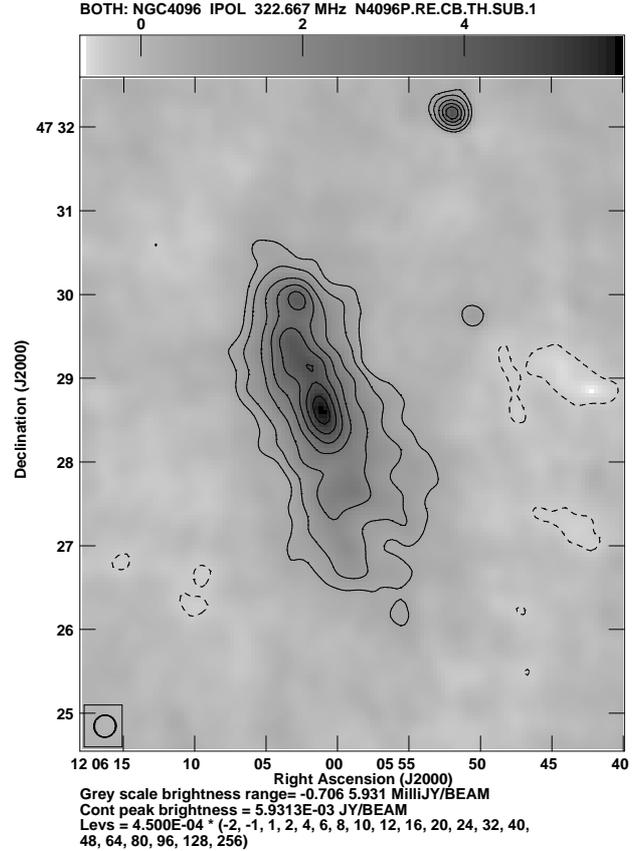}
\caption{Thermal emission subtracted 0.33 GHz map (in contour \& Gray scale) of NGC 4096.
Resolution is 15.9$^{''}\times 15.3^{''}$.}
\end{figure}.

\begin{figure}
\includegraphics[width=\columnwidth]{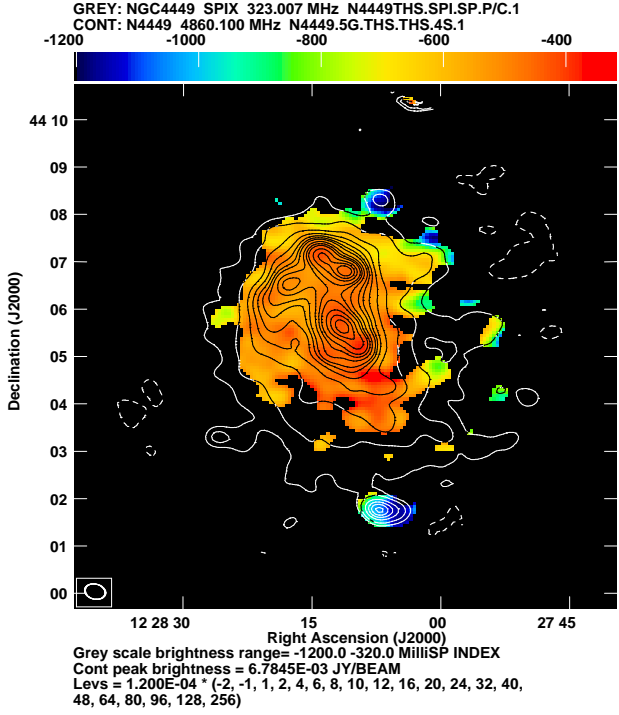}
\caption{Thermal emission subtracted 5 GHz map (contour) and non-thermal
spectral index of NGC 4449 (colour) between 0.33 and 5 GHz. Resolution is 26.1$^{''}\times 19.1^{''}$.}
\end{figure}

\begin{figure}
   \includegraphics[width=\columnwidth]{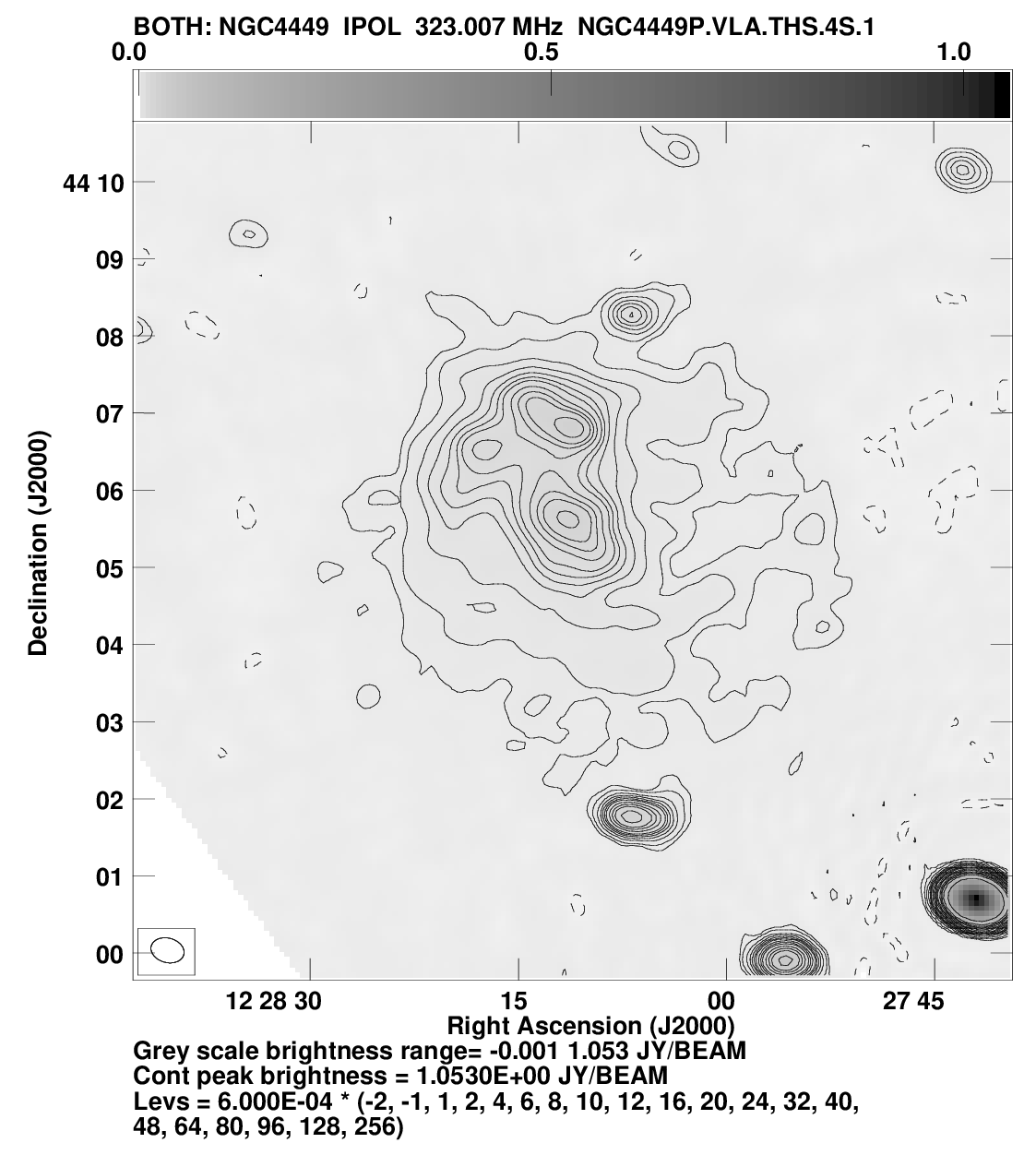}
\caption{Thermal emission subtracted 0.33 GHz map (in contour \& Gray scale) of NGC 4449.
Resolution is 26.1$^{''}\times 19.1^{''}$.}
\end{figure}.

\begin{figure}
\includegraphics[width=\columnwidth]{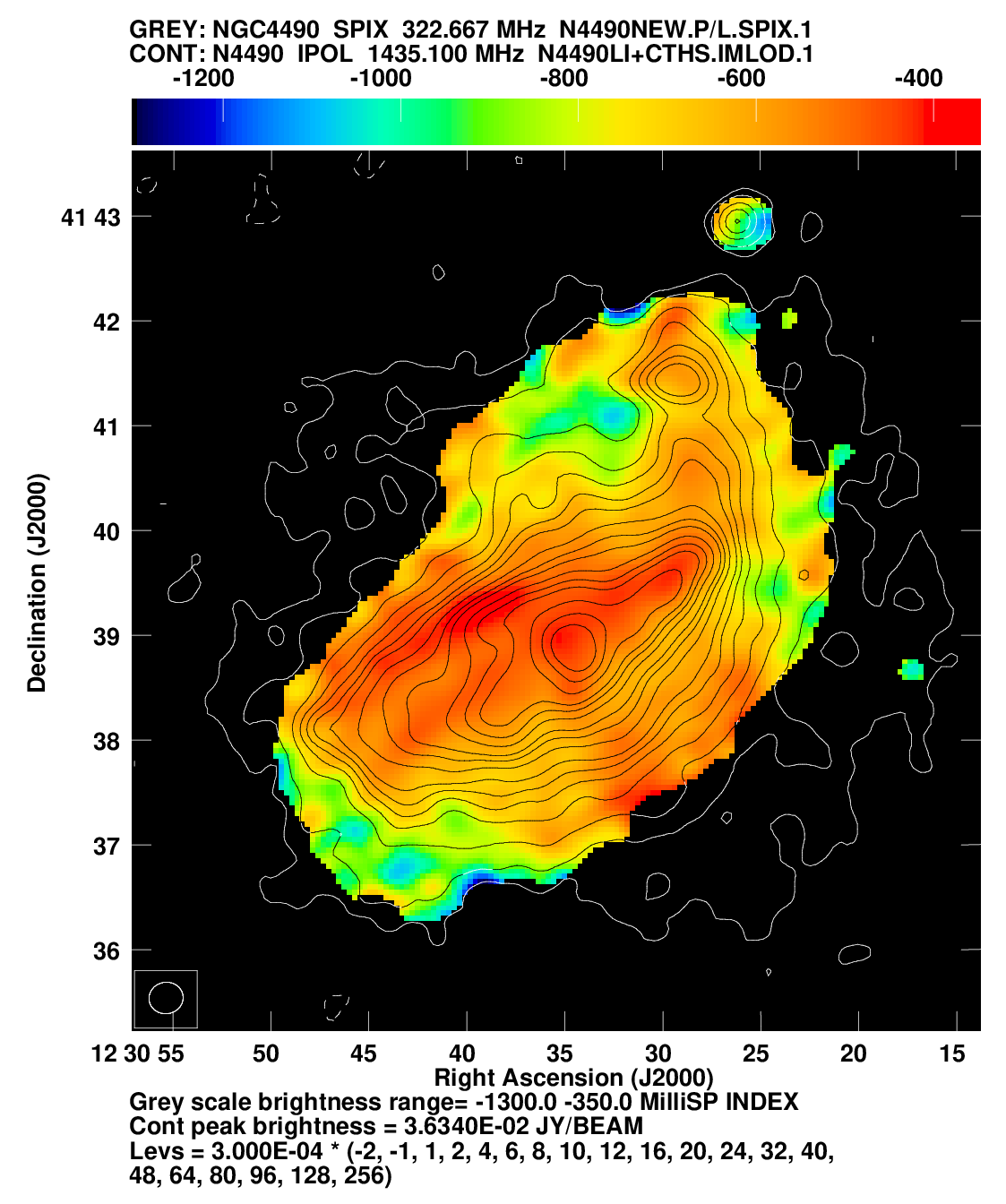}
\caption{Thermal emission subtracted 1.4 GHz image (contour) and non-thermal spectral index of NGC 4490 (colour) between 0.33 and 1.4 GHz. Resolution is 19$^{''}\times 18^{''}$.}
\end{figure}

\begin{figure}
\includegraphics[width=\columnwidth]{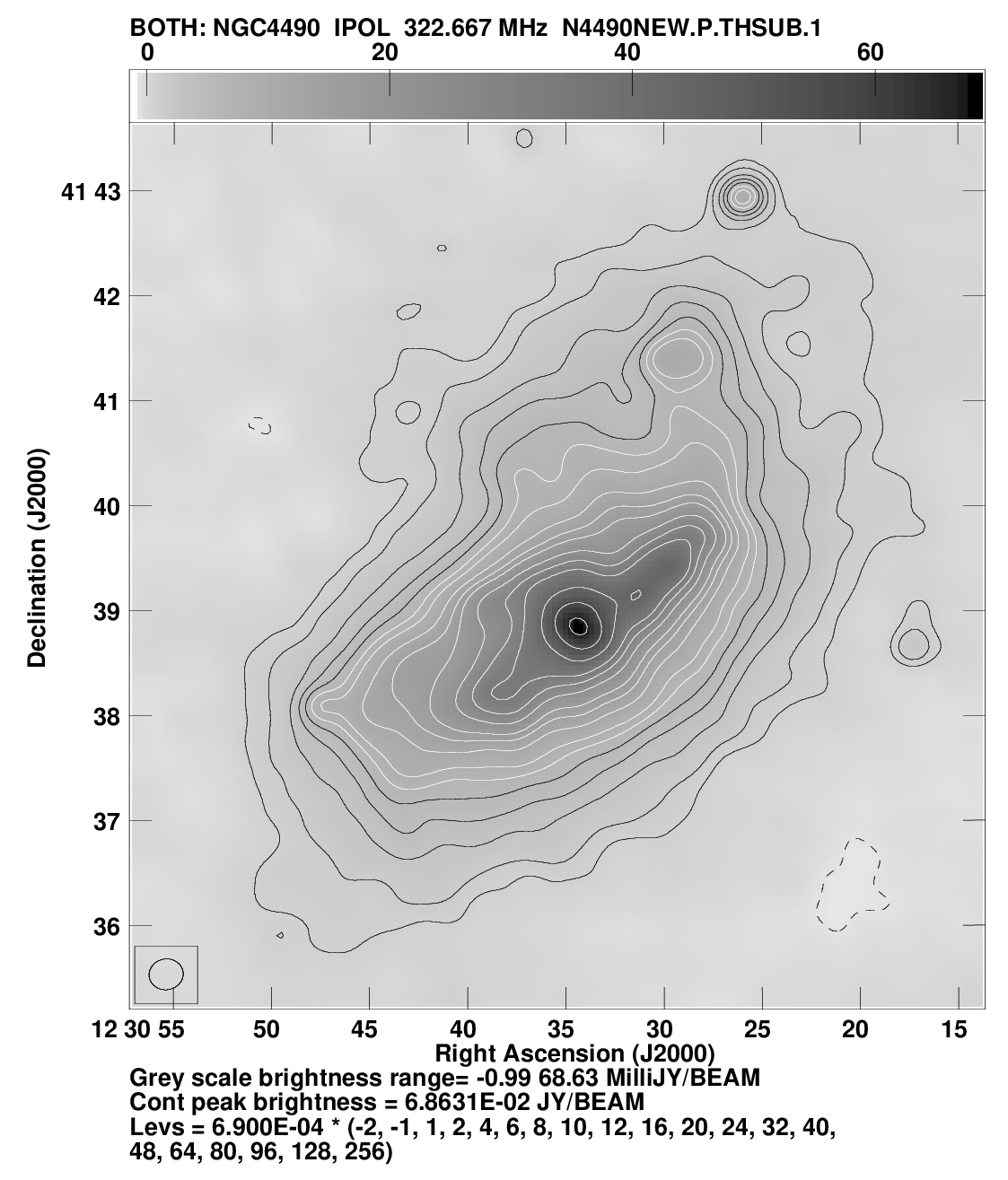}
\caption{Thermal emission subtracted 0.33 GHz map (in contour \& Gray scale) of NGC 4490.
Resolution is 19$^{''}\times 18^{''}$.}
\end{figure}.

\begin{figure}
\includegraphics[width=\columnwidth]{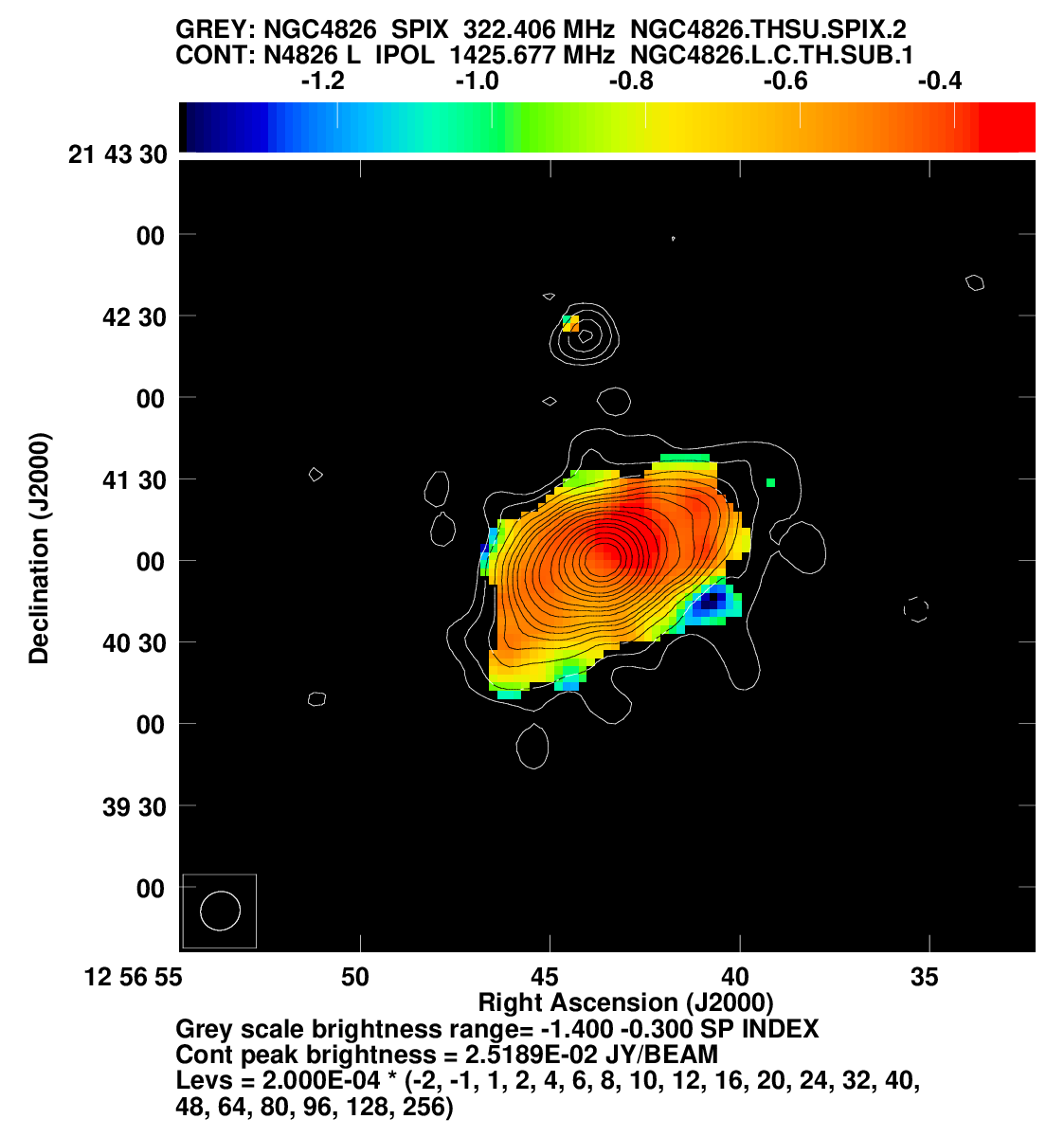}
\caption{Thermal emission subtracted 1.4 GHz image (contour) and non-thermal
spectral index of NGC 4826 (colour) between 0.33 and 1.4 GHz. Resolution is 15$^{''}\times 14^{''}$.}
\end{figure}

\begin{figure}
\includegraphics[width=\columnwidth]{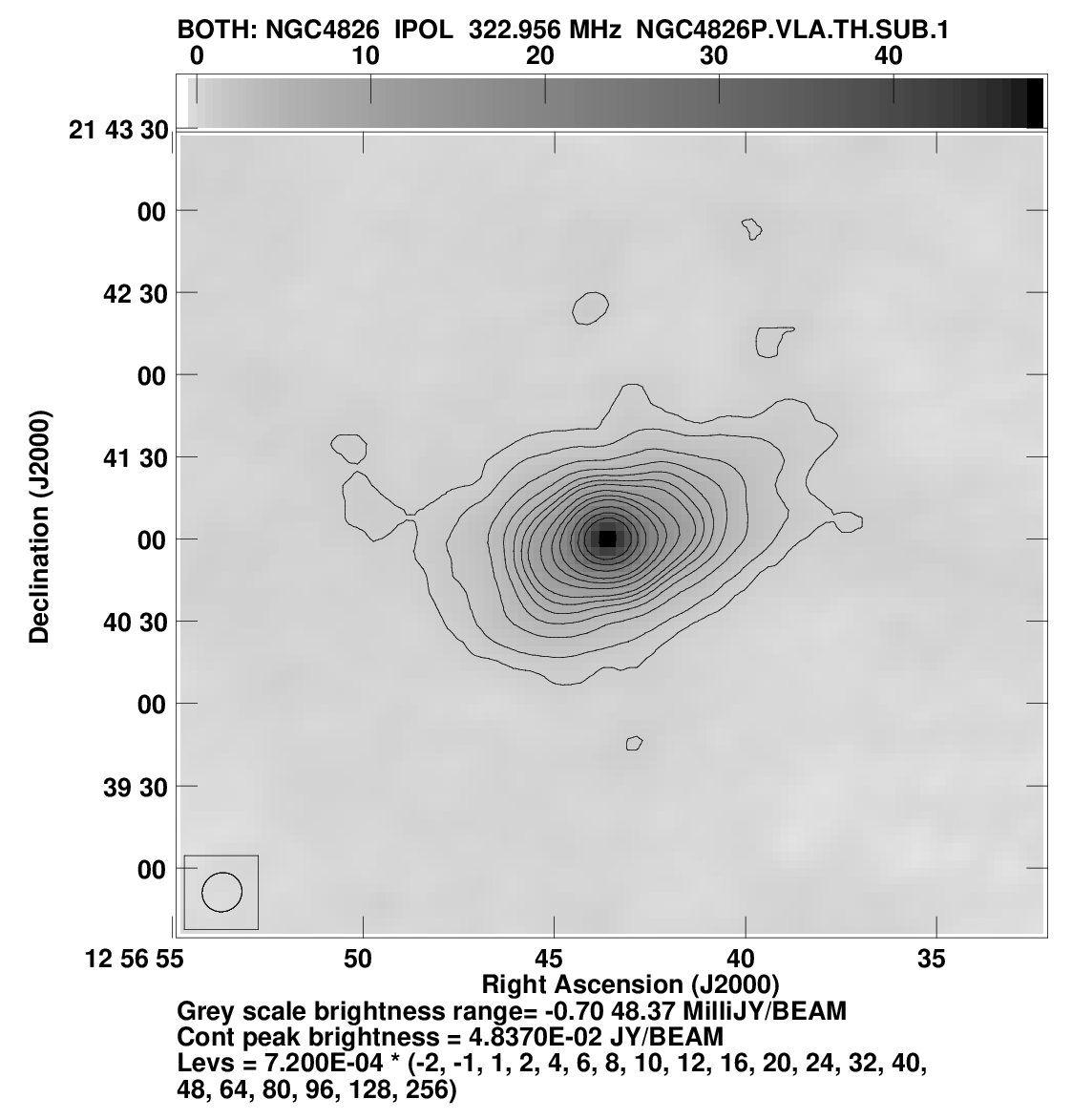}
\caption{Thermal emission subtracted 0.33 GHz map (in contour \& Gray scale) of NGC 4826.
Resolution is 15$^{''}\times 14^{''}$.}
\end{figure}.

\begin{figure}
\includegraphics[width=\columnwidth]{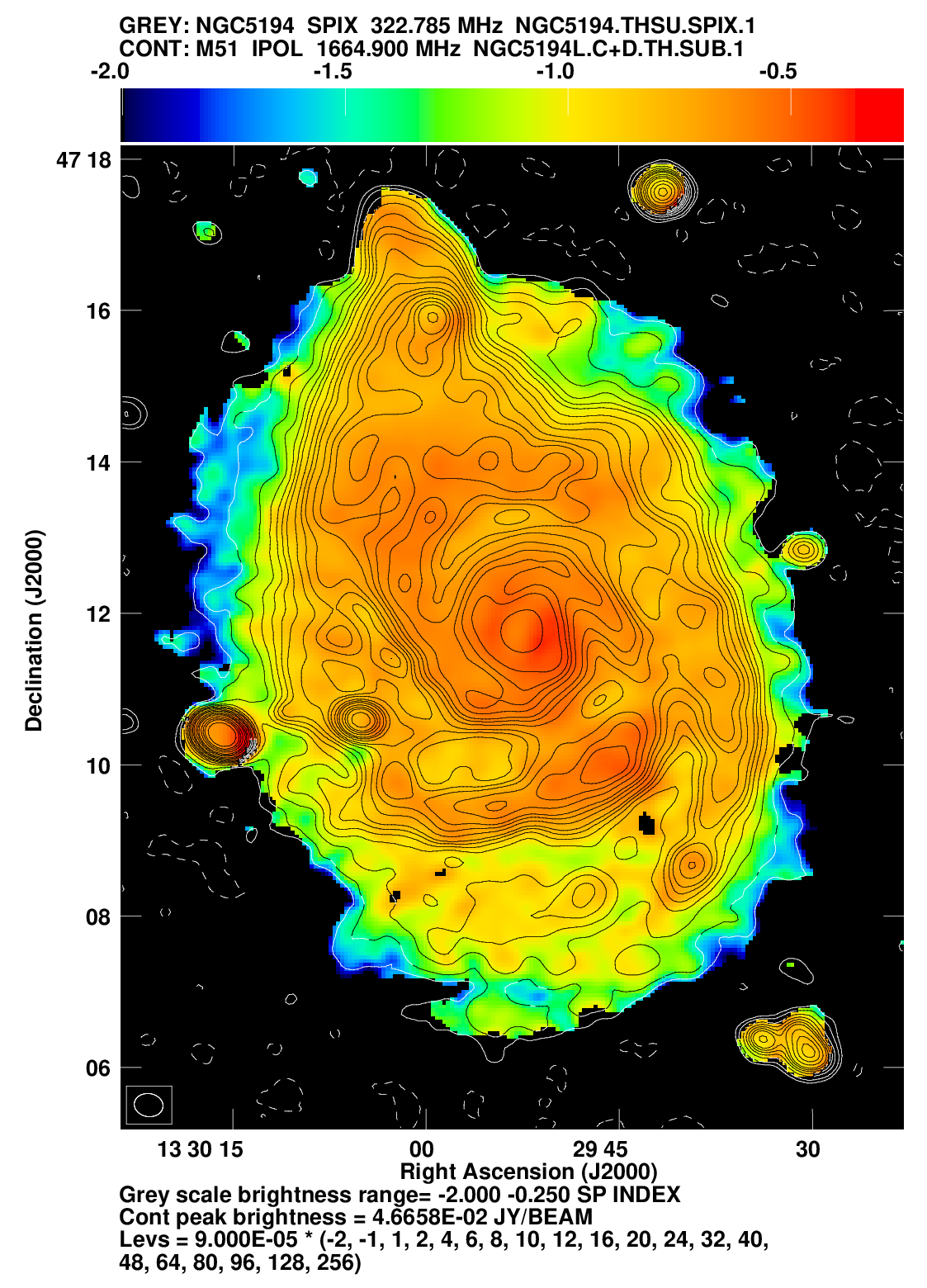}
\caption{Thermal emission subtracted 1.4 GHz image (contour) and non-thermal
spectral index of NGC 5194 (colour) between 0.33 and 1.4 GHz. Resolution is 23$^{''}\times 18^{''}$.}
\end{figure}

\begin{figure}
   \includegraphics[width=\columnwidth]{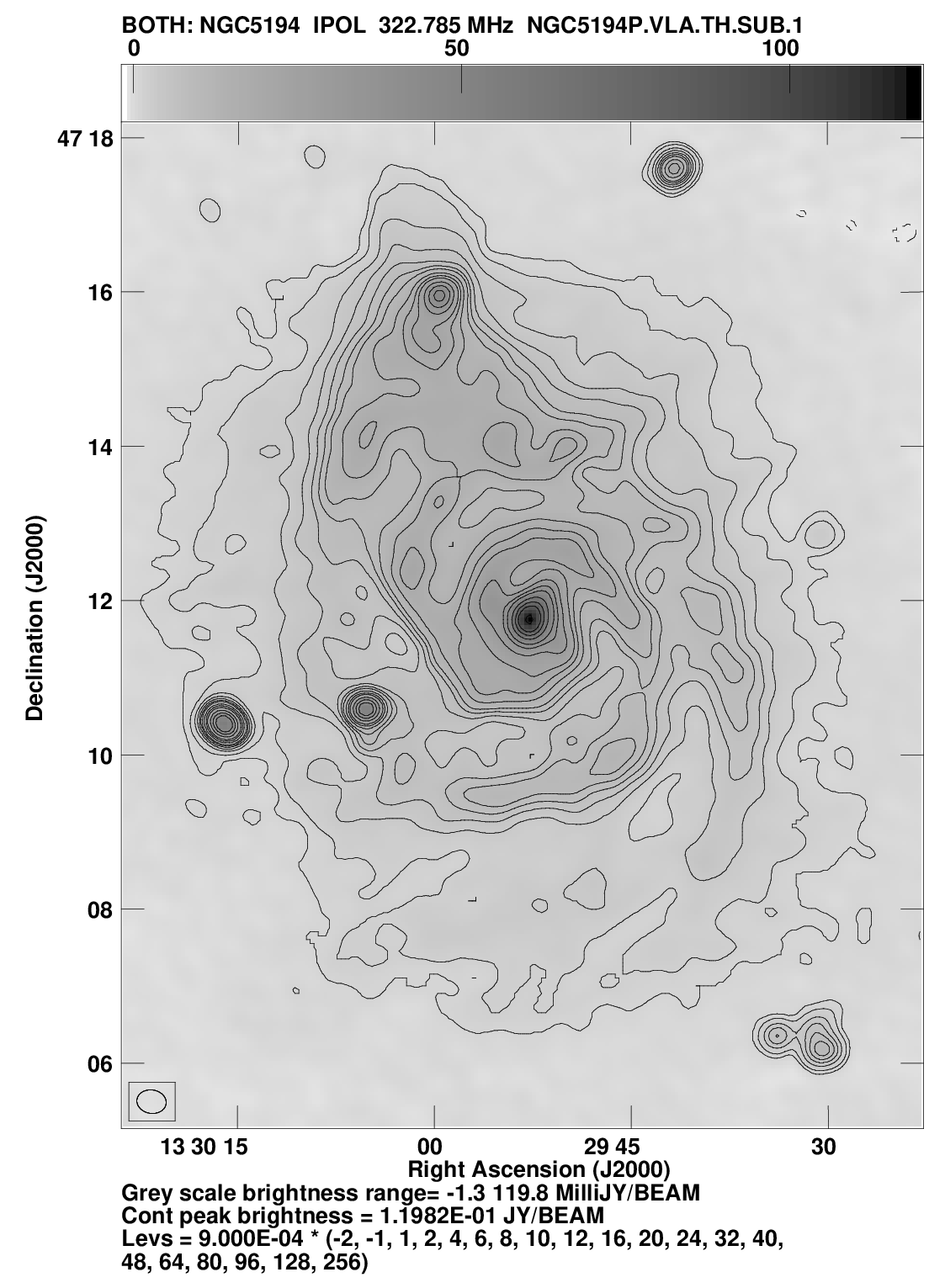}
\caption{Thermal emission subtracted 0.33 GHz map (in contour \& Gray scale) of NGC 5194.
Resolution is 23$^{''}\times 18^{''}$.}
\end{figure}

\subsection{Spectral index}

As mentioned in Sect. 1, non-thermal emission has a spectrum close to $- 0.5$
at their origins in star forming regions where thermal fraction is high. 
Several compact emissions are seen within NGC 2683, 3627 and 4096, and their
spectral index maps show these emissions to have flatter spectrum
indicating them to be star forming sites. The spectrum of all the galaxies
gradually steepens as a function of distance from their centre (where thermal
fraction is high) due to propagation loss as they move away to periphery of the
galaxies, where thermal
fraction is lowest. To investigate this steepening
of spectrum from centre to the periphery of the galaxies, we present the
azimuthally averaged nonthermal spectrum of the galaxies as a function of galactocentric
distance in Figs.~23 to 29. This averaging was performed by the Aips task IRING
using the known position and inclination angle of the galaxies, which are also
shown in Table-2. We have taken distances of these galaxies from
\citet{Dale2009ApJ}, which are shown in Table-2.

\begin{figure}
\includegraphics[width=\columnwidth]{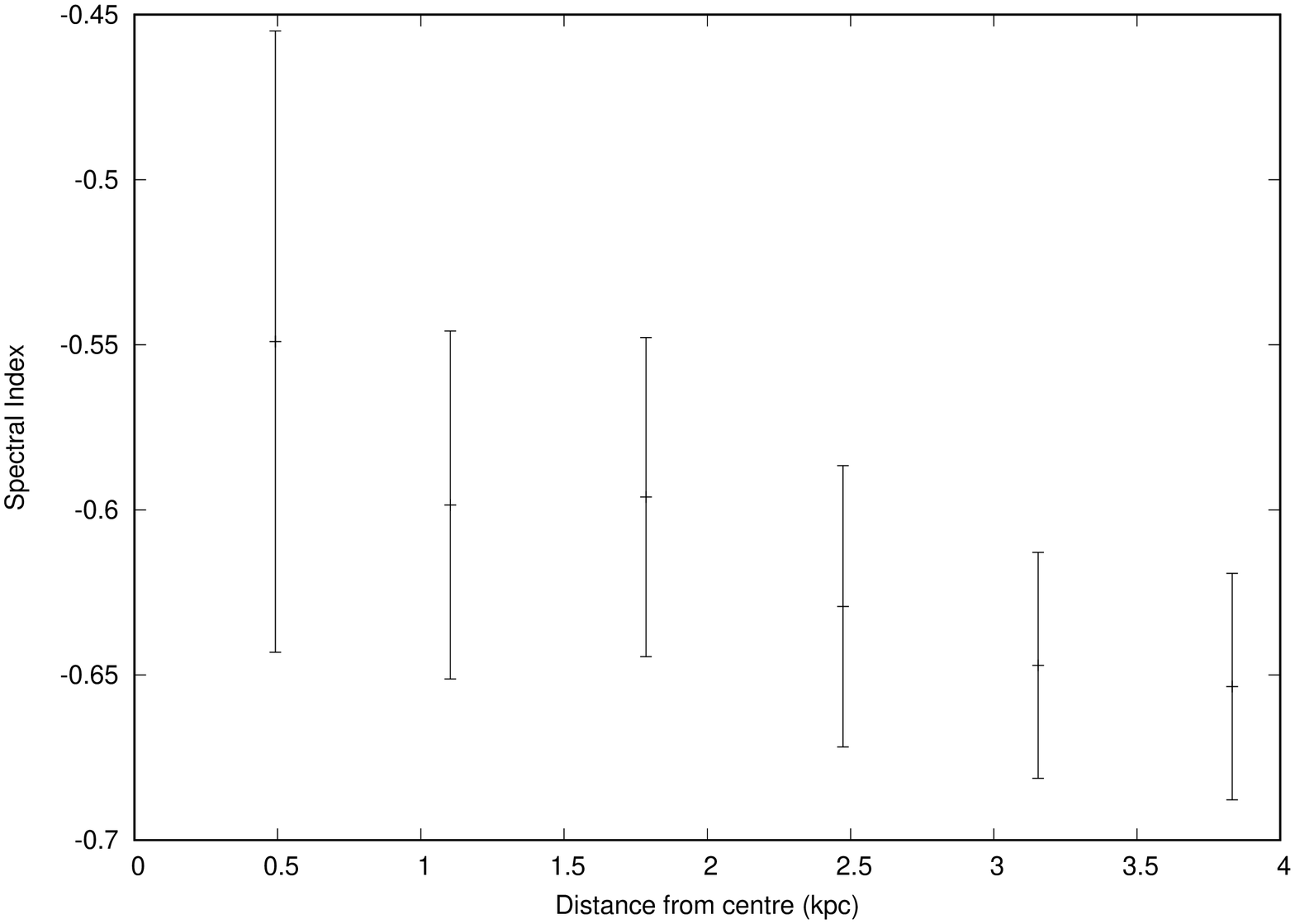}
\caption{Radial variation of non-thermal spectral index for NGC 2683.}
\end{figure}

\begin{figure}
\includegraphics[width=\columnwidth]{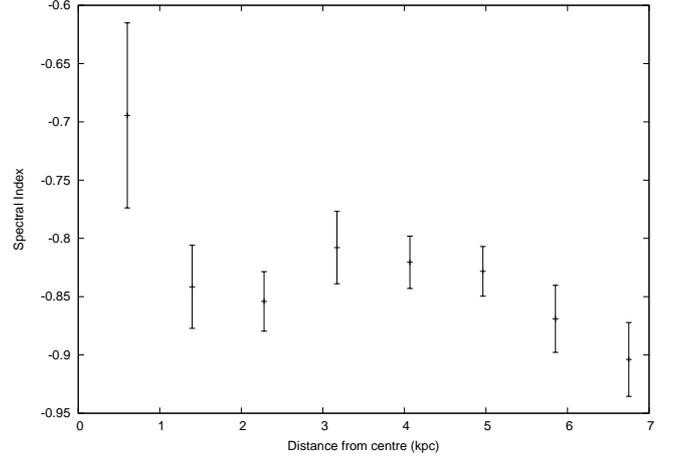}
\caption{Radial variation of non-thermal spectral index for NGC 3627.}
\end{figure}

\begin{figure}
   \includegraphics[width=\columnwidth]{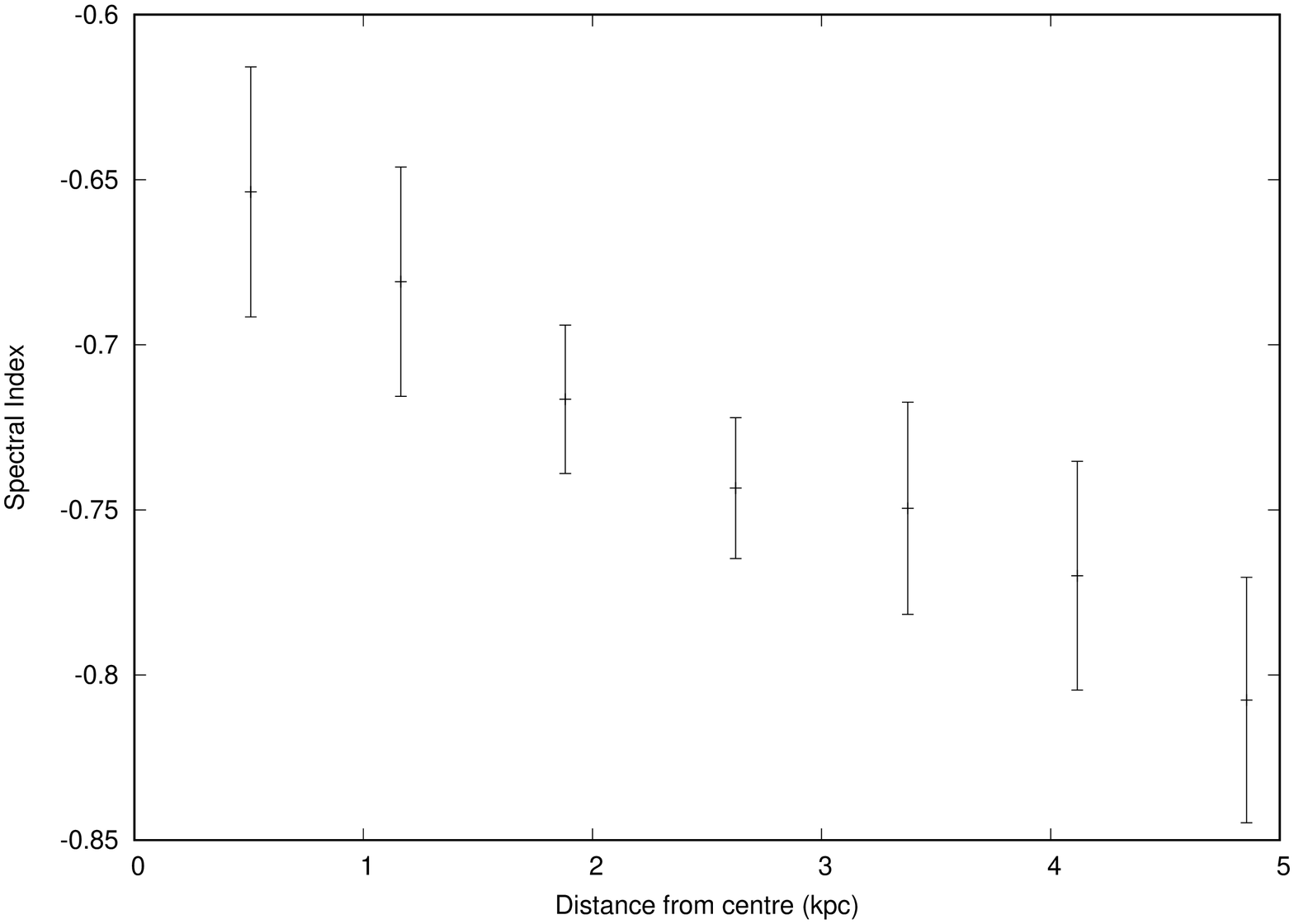}
\caption{Radial variation of non-thermal spectral index for NGC 4096.}
\end{figure}

\begin{figure}
   \includegraphics[width=\columnwidth]{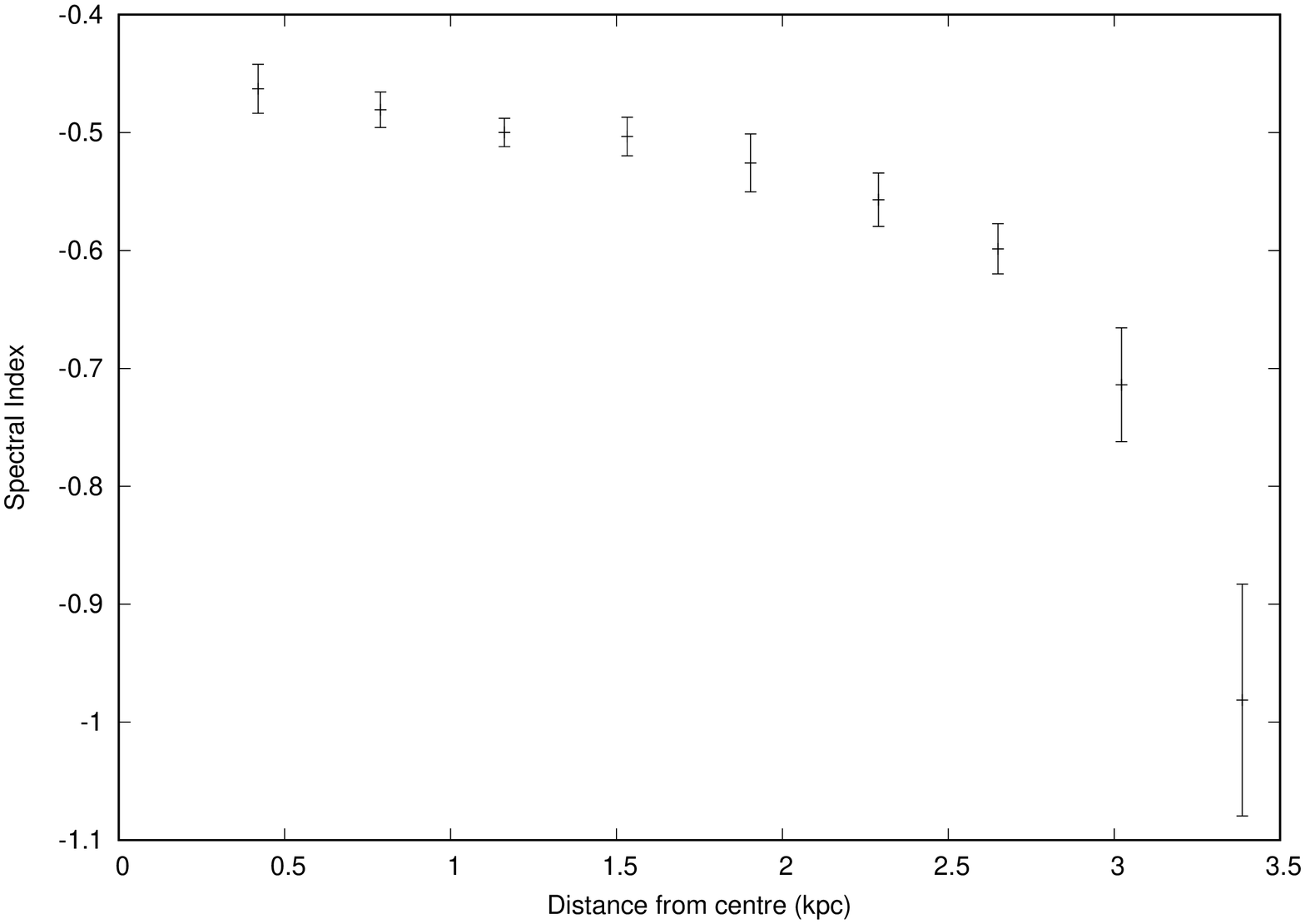}
\caption{Radial variation of non-thermal  spectral index for NGC 4449.}
\end{figure}

\begin{figure}
\includegraphics[width=\columnwidth]{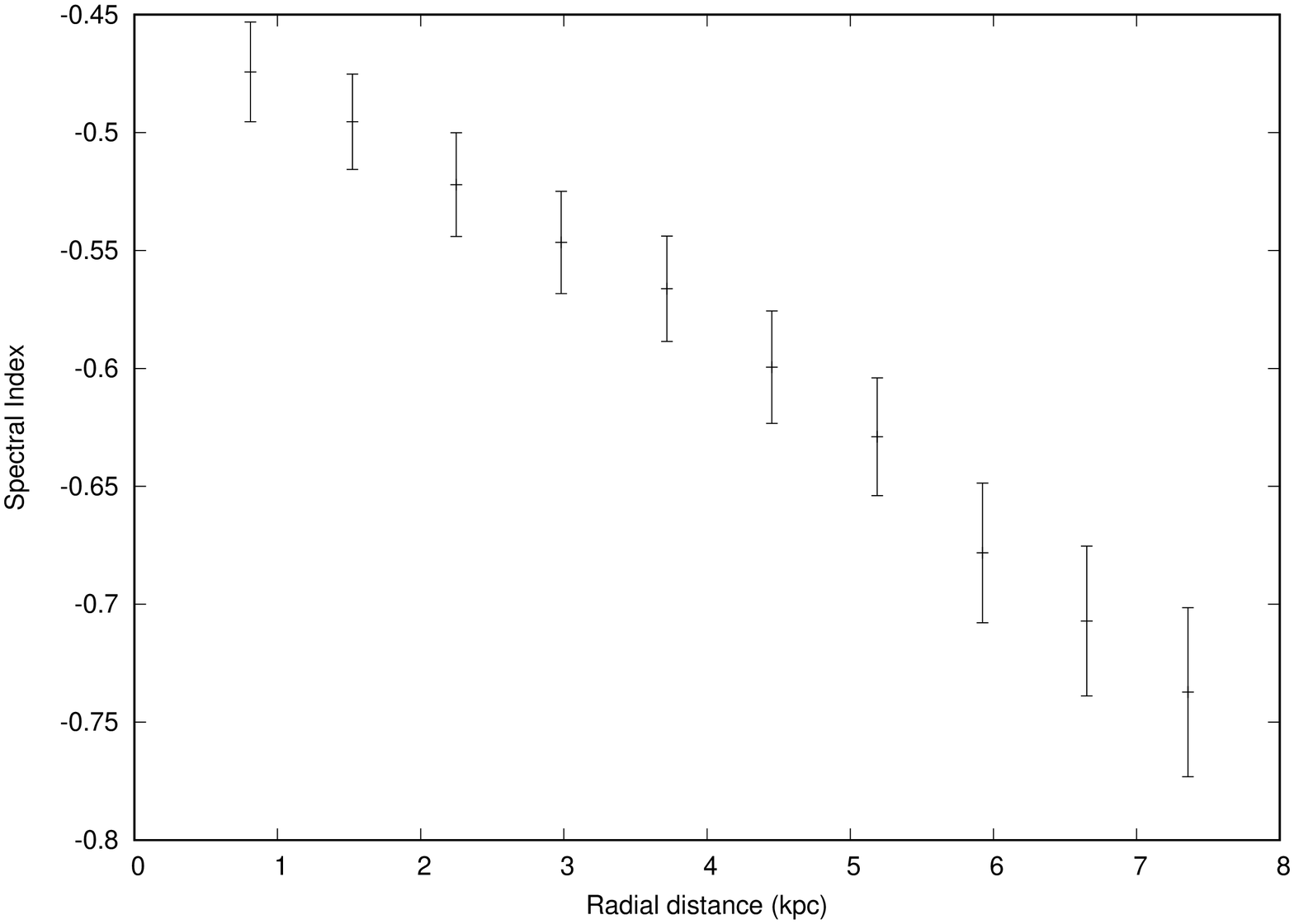}
\caption{Radial variation of non-thermal spectral index for NGC 4490.}
\end{figure}

\begin{figure}
   \includegraphics[width=\columnwidth]{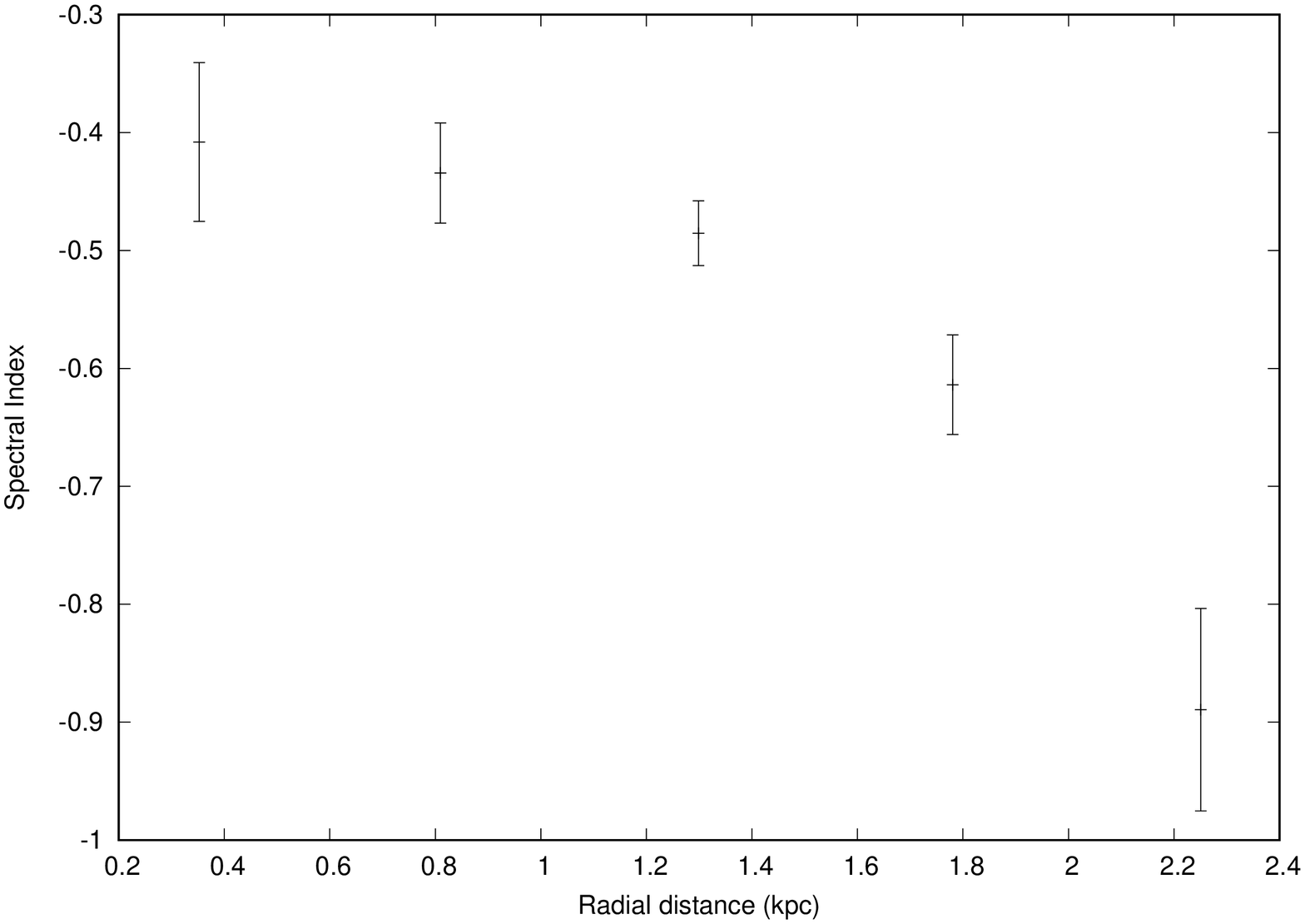}
\caption{Radial variation of non-thermal  spectral index for NGC 4826.}
\end{figure}

\begin{figure}
\includegraphics[width=\columnwidth]{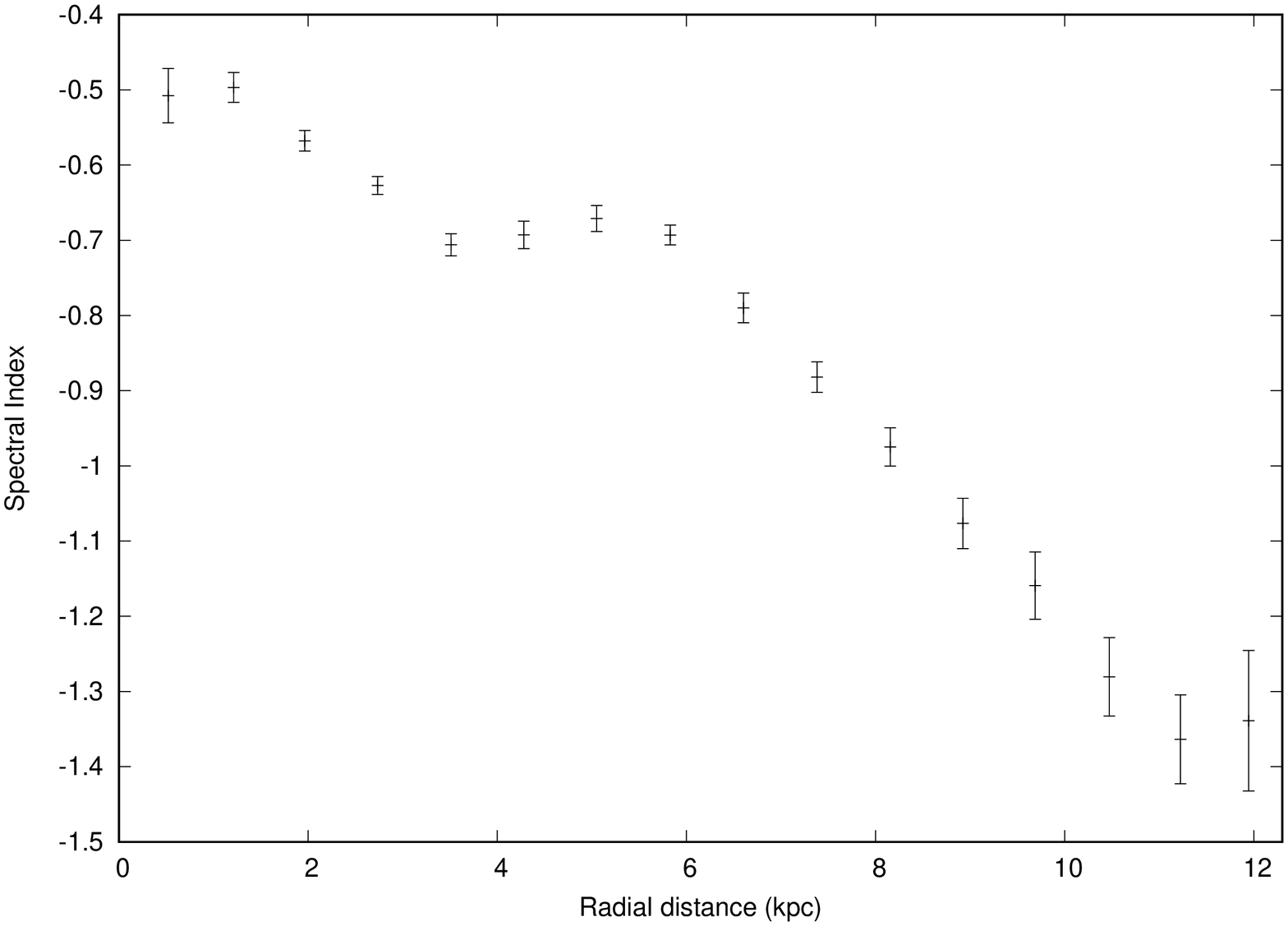}
\caption{Radial variation of non-thermal spectral index for NGC 5194.}
\end{figure}

For all the galaxies steepening of the spectrum from the
centre to the periphery is observed. 
It should be noted that star formation happens all across the disk of galaxies
(especially in arms of spirals) and radial averaging as done above would
average the flatter spectrum of star forming regions at different
galactocentric distances with the steeper spectrum of non star-forming region
at the same radius (with different position angle).

\subsection{Cosmic ray propagation within the galaxies}

Mean free path of UV photons ($\sim$100 pc) that heat the dusts in galaxies is
much less than that of CRes at $\sim 1$ GHz ($\sim1-2$ kpc) \citep{Murphy2006ApJ}.
CRes are believed to move far away from the source where diffusion plays a
significant role \citep{Lerche1982A&A,Ptuskin2001}. 
If we consider propagation only through diffusion, assume isotropic flow of CRes
from a compact source, constant spectral index of synchrotron emission and
constant magnetic field, then the emission is really dependent on the
densities of the CRes as they propagate from the sources.
Then as shown in Appendix, after a time (t) from turning-on of a CR source, CRe density
would follow a Gaussian distribution with distance from source. 
FWHM of Gaussian profile is determined by the denominator of the exponential,
which is $\sqrt{D\tau_{\nu}}$.
The above simple model of CRe diffusion would therefore cause the radio image
of a galaxy to appear as a convolution with a Gaussian of its FIR image
(consisting of many compact and extended sources).  In practice, this would cause smearing of
high frequency images of a galaxy in radio band \citep{Bicay&Helou1990ApJ}. 
Considering most of the emission occurs at critical frequency ($\nu_c$), for
synchrotron emission at a frequency $\nu$, the half life of synchrotron
emission ($T_{(1 / 2)\nu}$) in a constant magnetic field is $\propto  \nu^{-
0.5}$.
Therefore, ratio of propagation length scales at those two frequencies will
be $(\nu_1 / \nu_2)^{- 1 / 4}$, which can be tested from our observations.
Numerical simulations
considering isotropic pitch angle distribution of CRes
with power law distribution of energy \citep{GINZBURG1965} having power law
index between 2 and 3 show that ratio of emission between time $T=0$ and after
$T_{(1 / 2)\nu}$ for both the frequencies between 0.33 GHz and 6 GHz remains about
the same.  As the fall in emission at the two frequencies after respective
$T_{(1 / 2)\nu}$ occur by the same factor, the above estimation of the ratio of
propagation scale at two different frequencies remains valid in general.
It should be noted that CRes can also propagate through other mechanisms, which
include galactic wind, streaming instability \citep{Amato2018} or a combination of
the above three mechanisms. Depending on the mechanisms responsible, the ratio of
propagation lengths for the two frequencies would differ. For CRe propagation
through galactic wind, the ratio is independent of frequency, and for
streaming instability (distance travelled $\propto \tau $) the ratio would be
$(\nu_1 / \nu_2)^{- 1 / 2}$ \citep{Tabatabaei2013}.  In this paper, we
investigate whether the above simple model of diffusion is consistent with our
observations.

To estimate the propagation distance of the CRes, we assume that the CRes loss
energy mostly by synchrotron emission such that the exponent of radio FIR correlation
is unity. Within galaxies this can happen when CRe propagation do
not play important role as in flat spectrum star forming regions \citep{Heesen2019}.
To attain the same from the available maps of the
galaxies, we use the method of \citet{Berkhuijsen2013MNRAS}.
This method uses convolution by different kernel sizes of the FIR maps of an
object till the exponent of radio and convolved FIR map becomes unity.  Radio-FIR
correlation often uses monochromatic FIR emission at 70~$\mu$m, which we also
use for the above purpose.  A similar approach has also been used recently
by \citet{Vollmer2020} to study CRe propagation mechanism using 6 and 20 cm
radio maps of spirals. We convolved the 70~$\mu m$ Spitzer FIR maps of the
galaxies with different Gaussian FWHM using Aips task CONVL. The standard
radio-FIR correlation of the form
\begin{equation}
  S_{\tmop{radio}} = a  . (S_{\tmop{FIR}})^b
\end{equation}
was used to determined $b$ for different values of FWHM. 
In practice, we consider the two maps to have matched when the derived value of
$b$ is consistent with unity within the fitting error. \\
\tmtextbf{Blanking}:

Not all the emission regions
in the maps of each of the galaxies could be well fitted by unique values of
`$a$' and `$b$'. In certain cases, this happened when there are compact
background radio sources (often AGNs) seen through the galaxies in the maps. In such
cases, the regions containing the background sources were blanked from the
corresponding radio map. This approach was needed for NGC 5194,
where 4 background sources in the field were blanked. Moreover, for this
field, the part of the map showing the interacting galaxy NGC 5195 was
blanked. The central region of the galaxy is very bright in radio suggesting
contribution from the central AGN. Therefore, we have blanked the central
small diameter source from the FIR map of NGC 5194. 
NGC 4449 
is a dwarf irregular type of
galaxy and has three distinct dense star forming regions (Fig.~5). Two of
these compact regions in its north show different exponents in the radio-FIR maps
than the rest of the galaxy. Therefore, to remove contribution from the
two star forming regions, we blanked them along with their surrounding pixels
(within a couple of beam-widths).
The only other galaxy that required blanking was
NGC 4490. As it is interacting with its smaller companion NGC 4485 towards its
North-East, the part containing the companion and the bridge of emission
joining it was blanked.\\
\tmtextbf{Correlation of the convolved FIR maps with radio:}

To get a reasonable fit of equation (1) to the pixel values of the above maps,
we have chosen cutoff in radio intensities of $\sim 2 - 5$ times than the
corresponding map rms. (We have set the lower cutoff high to better reject
noisy data.  However, to get a fit over a large fraction of data for certain
galaxies with poorer signal to noise ratio, the cutoff was brought down up to 2
$\sigma$).
In the radio-FIR plots, multiple radio flux
densities (along y-axis) do occur within very close range of FIR (along x-axis).
Removing lower values along y-axis for a particular point along x-axis
would bias the fit. Therefore, when cut-offs along y-axis were applied, 
we also applied a corresponding lower cut-off along x-axis such that most of the y-axis
points below the threshold gets excluded from the fit due to cut along x-axis.
Near the highest flux densities in the
plots, systematic deviations were noted for
several galaxies. In such cases, upper cutoffs were also applied while
ensuring that it does not cause rejection of more than a few percent of the total pixels.
The fitted values of convolution FWHM corresponding to 0.33, 1.4 and $\sim$6 GHz
maps are shown in Table-3. 

The following procedure was used to estimate FWHM of the convolving Gaussian.
We initially convolved the 70~$\mu$m map of each galaxy with FWHM of
$1'$, $2'$ and $3'$ and compared the `$b$' values
for the three different propagation ranges. Depending on these comparisons
we further fine-tuned FWHM of the kernels so that `$b$' reaches
unity for 0.33, 1.4 and $\sim$5 GHz maps. The fits for each of the galaxies are
described below. \\
(i) NGC 2683: For this galaxy, the lower
cutoff for the fit to the 0.33 GHz radio data was 0.4 mJy/beam (2-$\sigma$).
For the reasons described above, FIR range for the fit was 3 to 6.8 Jy/beam.
Convolution with Gaussian FWHM of 115$^{''}$ produced correlation of 0.93
with 0.33 GHz radio map, with $\tmop{db} / d \theta =
0.3$ arc-min$^{- 1}$. Gaussian convolution FWHM for $b=1$ with the 330 MHz map is  $130'' \pm 24''$.
With radio map at 1.4 GHz, the lower limit of the radio flux
density was 0.2 mJy/beam, and FIR range between 0.3 to 1.7 Jy/beam. 
The size of Gaussian FWHM for $b$ to be unity was $21''\pm 8''$. 
With thermal subtracted radio map at 6 GHz, the lower limit of flux density was
0.055 mJy/beam, and FIR range 0.1 to 0.5 Jy/beam. For $b$ to be unity,
convolution FWHM is  $21'' \pm 7''$.

(ii) NGC 3627:  The
lower cutoff for the fit to the 0.33 GHz radio data was 3.0 mJy/beam
($\sim4-\sigma$). FIR range used for the fit was 7 to 22 Jy/beam. 
For a FWHM of $120''$, the fitted value of $b$ was 1.0$\pm$0.1. For $b$ to be
unity, convolution FWHM is $120'' \pm 24''$.
The 1.4 GHz image from VLA archival data has less structural sensitivity than
the 0.33 GHz map and we do not see the extended halo seen in the 0.33 GHz map
around the arms of the galaxy.. Therefore, we do not estimate the CRe
propagation FWHM from the 1.4 GHz observations.
We used its 4.8 GHz map made by combining single dish Effelsberg image with VLA
image \citep{SOIDA2001}. After subtracting the expected thermal emission, we
used a fit limit on FIR data between 1.5 and 8 Jy/beam and radio data above 0.4
mJy/beam. For $b$ to be unity, convolution FWHM is  $20'' \pm 4''$.

(iii) NGC 4096: The lower cutoff for the fit to the 0.33 GHz radio data was
0.5 mJy/beam (3-$\sigma$). FIR limits were 0.5 to 2.5 Jy/beam. With convolving
Gaussian FWHM size of $53''$, the fitted value of $b$ was 1.0$\pm$0.1,
yielding convolution FWHM of $53'' \pm 15''$. The same for the 1.4 GHz map is
$31'' \pm 10''$. Lower limit used for the fit from 1.4 GHz radio was 0.125
mJy/beam and FIR ranges for the fit were 0.15 to 1.7 Jy/beam. With 6 GHz radio
map, lower limit of the fit was 0.05 mJy/beam, and FIR limits used were 0.2 to
1.2 Jy/beam. However, there were certain inconsistencies in convolution sizes 
using the FIR map for $b$ to be unity. We could only get an upper limit on the
convolution size of 32$''$.

(iv) NGC 4449: 
With radio data between 0.6 to 5 mJy/beam and a lower limit of 3 Jy/beam in
FIR, the derived value of convolution FWHM is $90'' \pm 12''$ for the 0.33 GHz map.
The same for the 1.4 GHz map is $115'' \pm 15''$. 
The lower limit of radio flux density was 0.5 mJy/beam ($\sim 3 \sigma$), and
to avoid systematic deviation, an upper cutoff of 14 mJy/beam was used. FIR
lower limit was 4 Jy/beam.  The 1.4 GHz image from VLA archival data has less structural
sensitivity than the 0.33 GHz map. 
Therefore, we used its 4.8 GHz map made by combining single dish Effelsberg
data with VLA \citep{Chyzy2000}. After subtracting the expected thermal emission,
we used a fit limit on FIR data between 1. and 7 Jy/beam and radio data above
0.1 mJy/beam. For $b$ to be unity, convolution FWHM is  $62'' \pm 10''$.

(v) NGC 4490: With radio data above 1
mJy/beam and a lower limit of 2.5 to 18 Jy/beam in FIR, the derived value of
Gaussian convolution FWHM is $86'' \pm 17''$ for the 0.33 GHz map. The same for
the 1.4 GHz map is $55'' \pm 8''$. 
The lower limit of radio flux densities used for the fit was 0.5 mJy/beam ($5
\sigma$). To avoid systematic deviation from the fit, FIR range used was 1.0 to
12 Jy/beam.

(vi) NGC 4826: Emission from this galaxy is dominated by the central
region, which is not well resolved with an angular resolution of $18''$. Diffuse
extended emission is seen around it.
The centrally peaked emission is seen in radio and IR (Fig.~7). However,
contribution from hidden Active Galactic Nuclei (AGN) at the centre cannot be
ruled out for this galaxy in radio band.  With 0.33 GHz radio data above 0.34
mJy/beam (4$\sigma$) and an FIR lower limit of 0.5 Jy/beam, the value of $b$
using 70 $\mu$m map without any convolution is greater than unity indicating
FIR emission is more diffuse than radio.
Therefore, CRe propagation scale cannot be determined by the
above method for this galaxy.  Hence, we have kept the Gaussian convolution
FWHM blank for this object in Table-3. 

(vii) NGC 5194: With radio data between 1.0 and 25 mJy/beam and FIR data above
4 Jy/beam, the derived value of convolving Gaussian is $150'' \pm 10''$ for the
0.33 GHz map. The same for the 1.4 GHz map is $75'' \pm 10''$.  The lower and
upper limit of radio flux densities used for the fit was 0.2 to 10 mJy/beam. To
avoid systematic deviation, a lower cutoff of 0.7 Jy/beam was used for FIR.
We also compared the propagation scale at 0.14 GHz from LOFAR map
\citep{Mulcahy2014A&A} after thermal subtraction. With radio data above 1
mJy/beam and FIR data between 8 to 45 Jy/beam, the derived value of FIR
convolution size for $b=1$ is 193$\pm$14$''$ (7.8$\pm$0.6 kpc). It is
a factor of two larger than what is found by \citet{Heesen2019} for this galaxy.
They used convolution of hybrid-SFR surface density map to obtain unit exponent
for SFR surface density from hybrid-SFR and 1.4 GHz radio.
We note that their method uses a combination of FUV and 24 $\mu$m data to
define the star forming disk, while the method used in this work uses 70 $\mu$m FIR.

\begin{table*}
\caption{Gaussian convolution FWHM derived from radio-FIR correlation for $b = 1$ }
 \centering 
  \begin{tabular} {||c c c c c c c c c||}
\hline
Galaxy name & \multicolumn{2}{c} {FWHM at 0.33 GHz} & \multicolumn{2}{c} {FWHM at 1.4 GHz} & FWHM ratio between & \multicolumn{2}{c} {FWHM at $\sim$6 GHz} & FWHM ratio between\\
    \  & arc-sec & kpc & arc-sec & kpc &  0.33 \& 1.4 GHz &  arc-sec & kpc & 0.33 \& $\sim$6 GHz\\
\hline
    NGC 2683 & 130$\pm$24  & 4.8$\pm$0.9 & 21$\pm$8  & 0.8$\pm$0.3 & 6.2$\pm$2.6 & 21$\pm$7 & 0.8$\pm$0.3 & 6.2$\pm$2.3 \\
    NGC 3627 & 120$\pm 24$ & 6.0$\pm$1.2 &     --    &      --&  --   &20$\pm$4 & 1.0$\pm$0.2 & 6.0$\pm$1.7 \\
    NGC 4096 & 53$\pm 15$ & 2.1$\pm$0.6 & 31$\pm$10\quad & 1.2$\pm$0.4 & 1.7$\pm$0.7 & $<32^{''}$ & $<1.3$ & $>$1.6\\
    NGC 4449 & 90$\pm$12 & 1.8$\pm$0.25 & --              &  --        & -- & 62$\pm$10 & 1.3$\pm$0.2 & 1.4$\pm$0.3\\
    NGC 4490 & 86$\pm$17 & 3.3$\pm$0.7 & 55$\pm$8  & 2.1$\pm$0.3  & 1.6$\pm$0.4 & --       &  -- &  -- \\
    NGC 4826 & -             &  -          &  -        & -  & -- & -- & -- & --\\
    NGC 5194 & 150$\pm$10 & 6.1$\pm$0.4 & 75$\pm$10 & 3.1$\pm$0.4 & 2.0$\pm$0.3 & -- & -- & --\\
\hline
  \end{tabular}
\end{table*}

\subsection{Spatially resolved radio-FIR correlation}

Spatially resolved study of radio-FIR correlation has been carried out using 1.4
and 0.33 GHz radio maps and 70 $\mu$m FIR map. 
These maps were blanked as described in Sect.~4.3.

Radio-FIR correlations were studied by fitting Equation.~1.
Arm and inter-arm region could be isolated in NGC 5194. For other
galaxies we could not separate arm and interarm region visually and used the
spatially resolved emission across the galaxies for this study.  
All images were convolved to the same resolution as shown in
Table-2 using the task CONVL in Aips. The cut-offs used for the correlation
fits in radio are the same as described in Sect.~4.3, where criteria used for
FIR ranges are also described.  Logarithm (base 10) of FIR ranges (in Jy/beam) used for
NGC 2683, 3627, 4096, 4449, 4490, 4826 with 0.33 GHz radio data were $-1 \, \text{to} \, -0.222$, 
$> -1$, $-1.3 \, \text{to} \, -0.22$, $>-1$, $> -1.33$, and $>$0.079 respectively.
FIR Ranges used (in dex) with 1.4 GHz radio data for the above galaxies were
$-.824 \, \text{to}\, -.222$, $>$0.079, $-1.398 \, \text{to}\, -0.22$, $>-1$, $> -1.22$ and $>-0.22$
respectively. The FIR ranges used for the arm regions of NGC 5194 were $>-0.7$
and $>-1.0$ with 0.33 and 1.4 GHz radio data respectively.  The above ranges
for interarm regions of NGC 5194 were $>-0.7$ and $>-1.3$ respectively.

The radio-FIR correlations for all the galaxies are presented in Figs.~30 to
37. In these plots, 0.5 dex has been added to the 0.33 GHz radio data to avoid
any overlap with the 1.4 GHz radio data.
All the correlation values are given in Table-4.
We have also provided the scatter of the correlation which is the rms
distance of all points from the fitted line. In all the figures mentioned above,
data points for the 1.4 GHz radio appear lower than the 0.33 GHz values.
All the plots except for NGC 5194 include all the pixels selected. However,
while estimating the errors, we have accounted for the number of pixels in a
synthesised beam (resolution), values of which are correlated. For NGC 5194, we
could separate arm and interarm regions visually using the Spitzer 24~$\mu$m map,
which we have selected by visual masking, and these plots are shown in Fig.~36
and 37. Exponents for the 0.33 GHz radio and FIR varies from 0.5 to even higher than
1. However, the exponent (b) using 1.4 GHz radio data is always higher than 0.33 GHz 
data except for NGC 4449. This is consistent with what was found by
\citet{Basu2012ApJ}.

\begin{figure}
\includegraphics[width=\columnwidth]{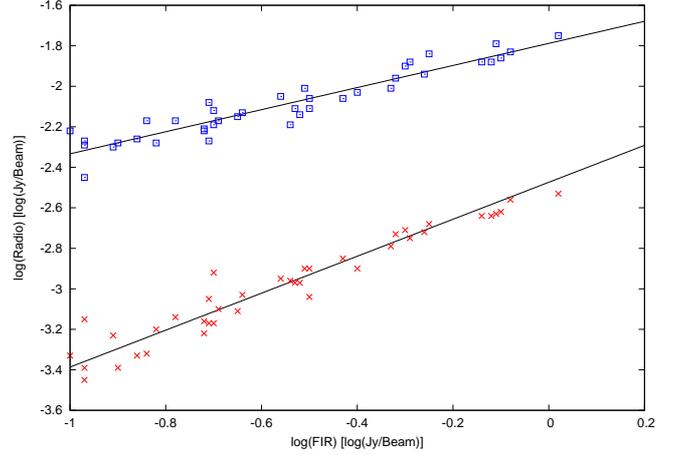}
\caption{70 $\mu$m FIR vs 0.33 and 1.4 GHz radio intensity (lower plot) of NGC 2683.
Note: To prevent overlap of 0.33 GHz data with the 1.4 GHz data on the plot, 
0.5 dex has been added to the 0.33 GHz data.}
\end{figure}

\begin{figure}
\includegraphics[width=\columnwidth]{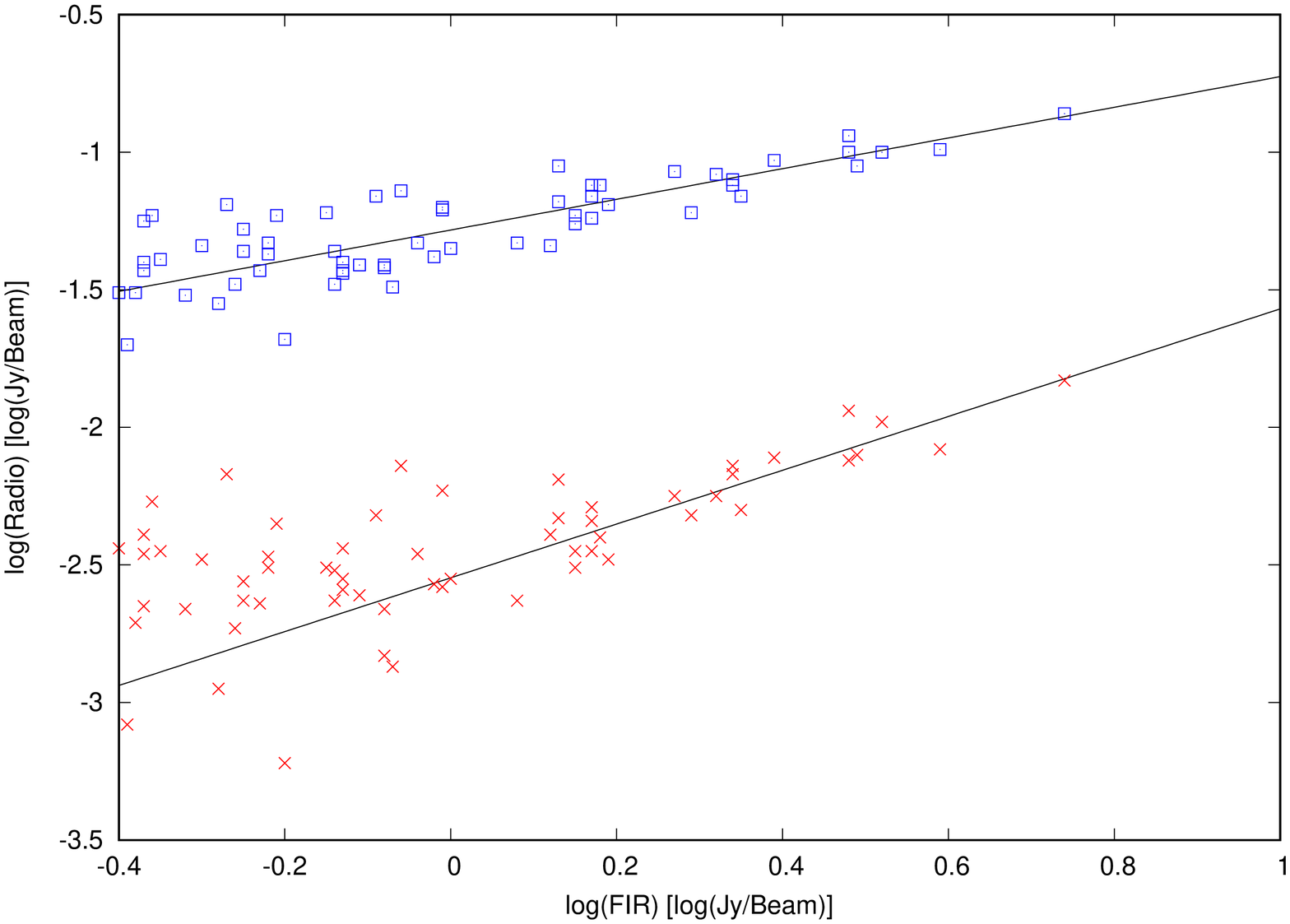}
\caption{70 $\mu$m FIR vs 0.33 and 1.4 GHz radio intensity (lower plot) of NGC 3627. 
To the 0.33 GHz data 0.5 dex has been added.}
\end{figure}

\begin{figure}
\includegraphics[width=\columnwidth]{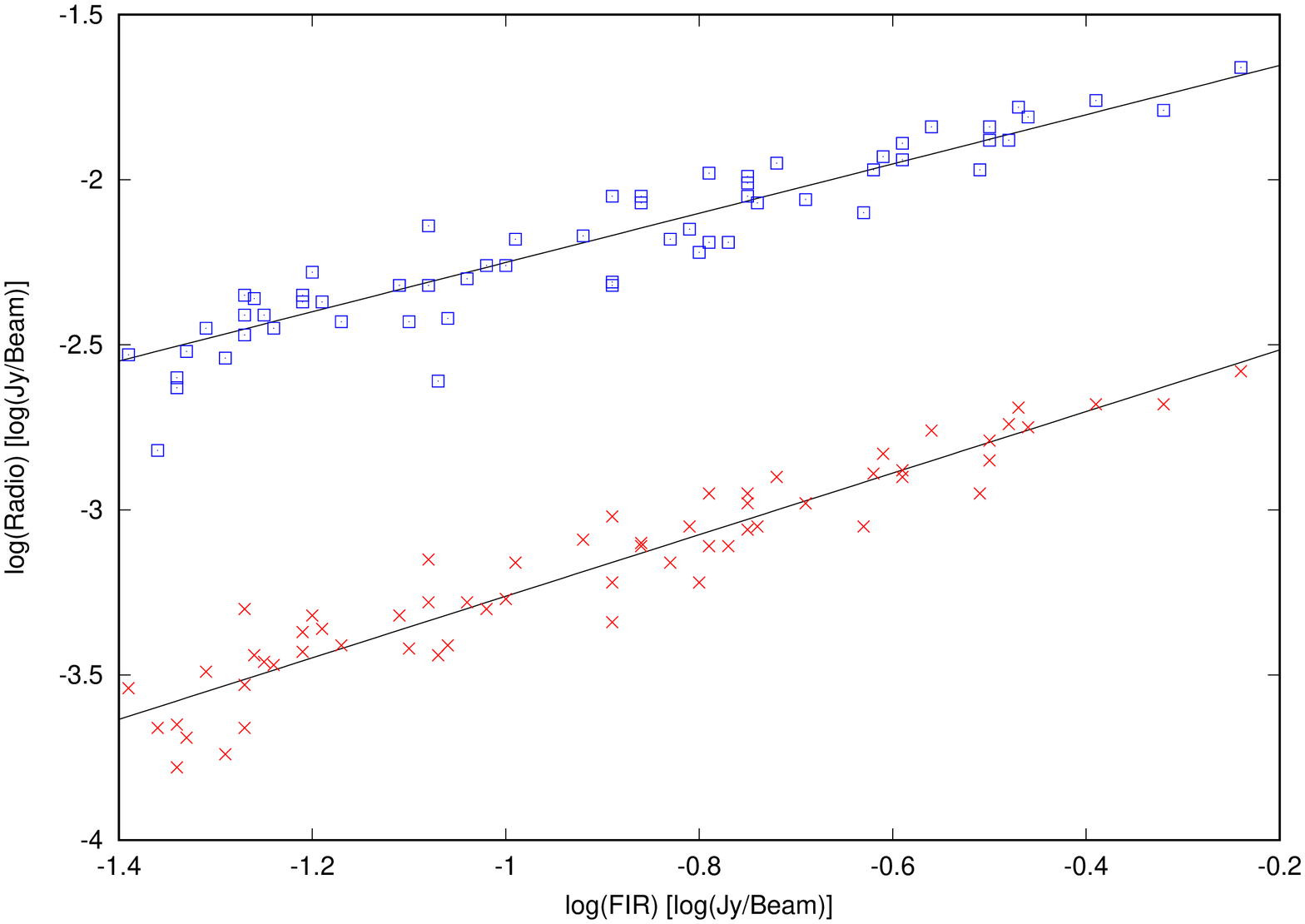}
\caption{70 $\mu$m FIR vs 0.33 and 1.4 GHz radio intensity (lower plot) of NGC 4096. 
To the 0.33 GHz data 0.5 dex has been added.}
\end{figure}

\begin{figure}
\includegraphics[width=\columnwidth]{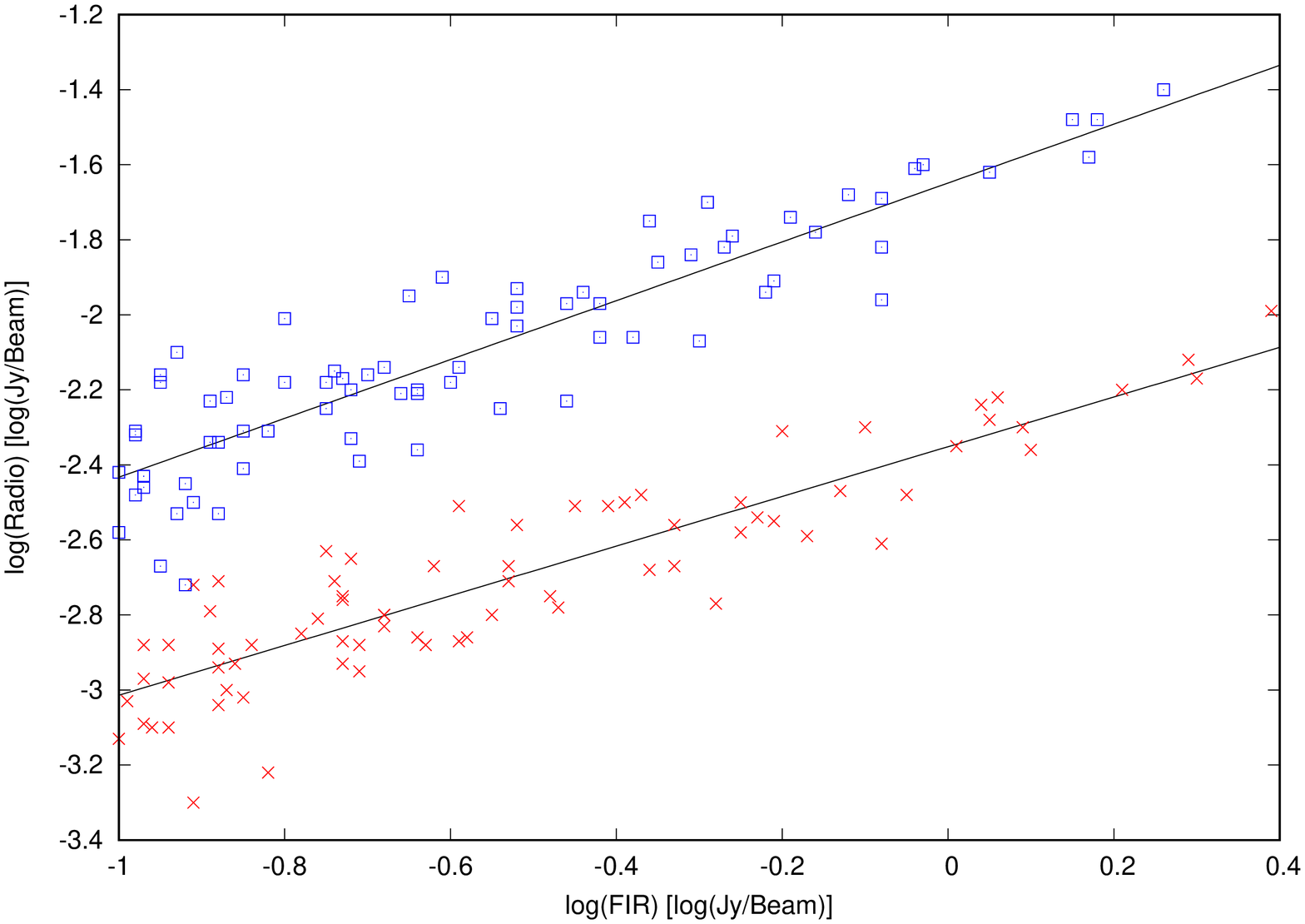}
\caption{70 $\mu$m FIR vs 0.33 and 1.4 GHz radio intensity (lower plot) of NGC 4449. 
To the 0.33 GHz data 0.5 dex has been added.}
\end{figure}

\begin{figure}
\includegraphics[width=\columnwidth]{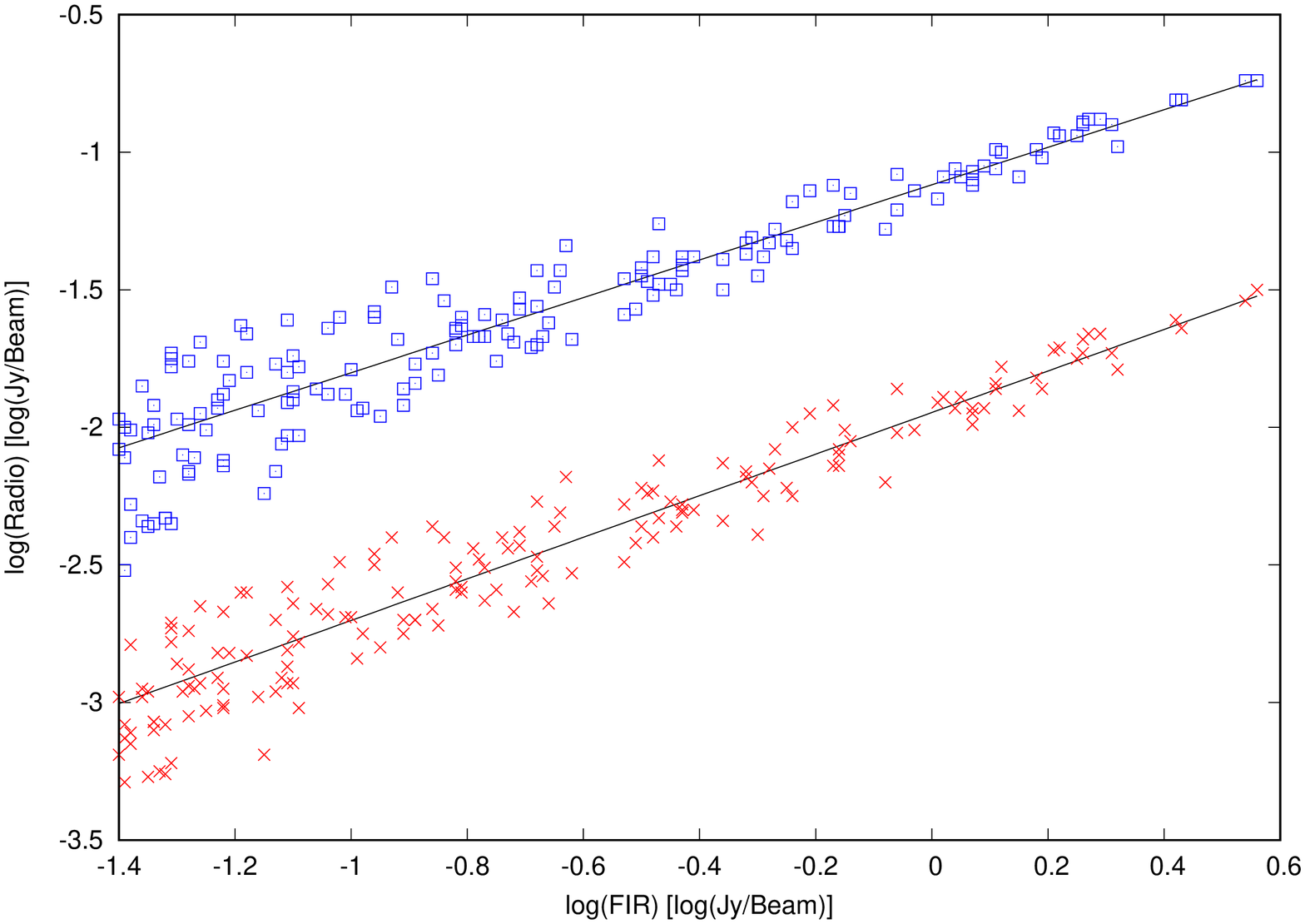}
\caption{70 $\mu$m FIR vs 0.33 and 1.4 GHz radio intensity (lower plot) of NGC 4490. 
To the 0.33 GHz data 0.5 dex has been added.}
\end{figure}

\begin{figure}
\includegraphics[width=\columnwidth]{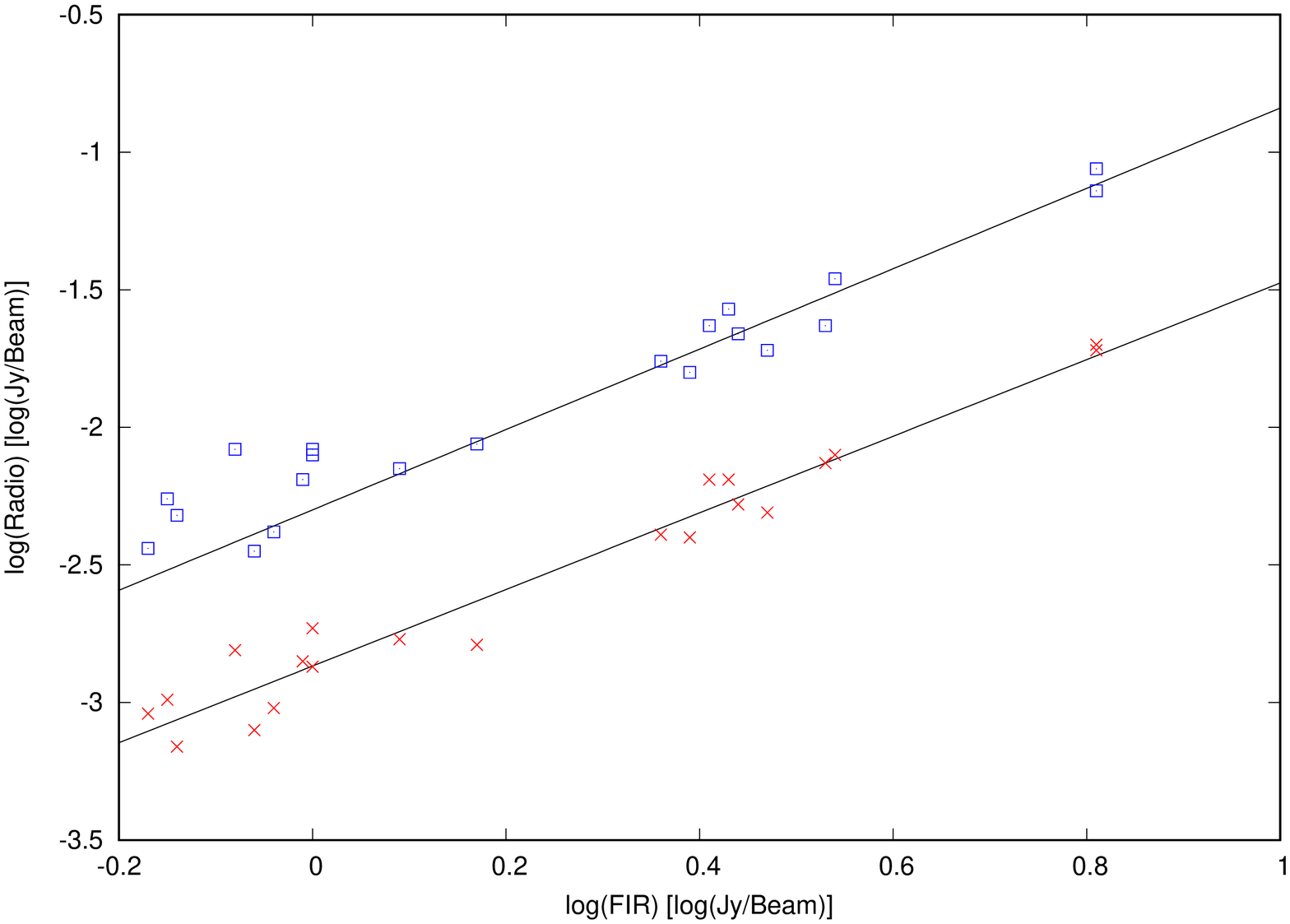}
\caption{70 $\mu$m FIR vs 0.33 and 1.4 GHz radio intensity (lower plot) of NGC 4826. 
To the 0.33 GHz data 0.5 dex has been added.}
\end{figure}

\begin{figure}
\includegraphics[width=\columnwidth]{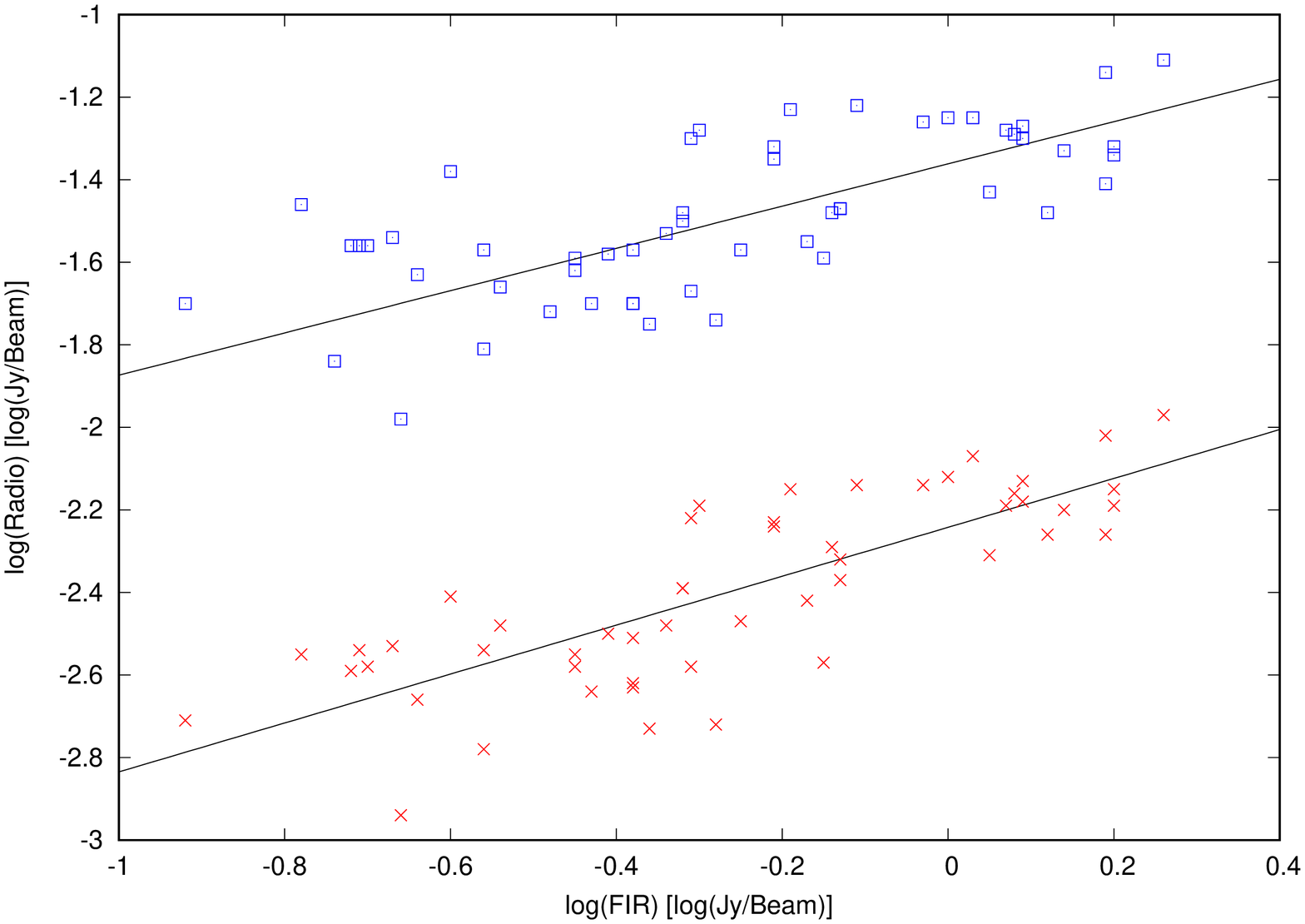}
\caption{70 $\mu$m FIR vs 0.33 and 1.4 GHz radio intensity (lower plot) in the arm region of NGC 5194.
To the 0.33 GHz data 0.5 dex has been added.}
\end{figure}

\begin{figure}
\includegraphics[width=\columnwidth]{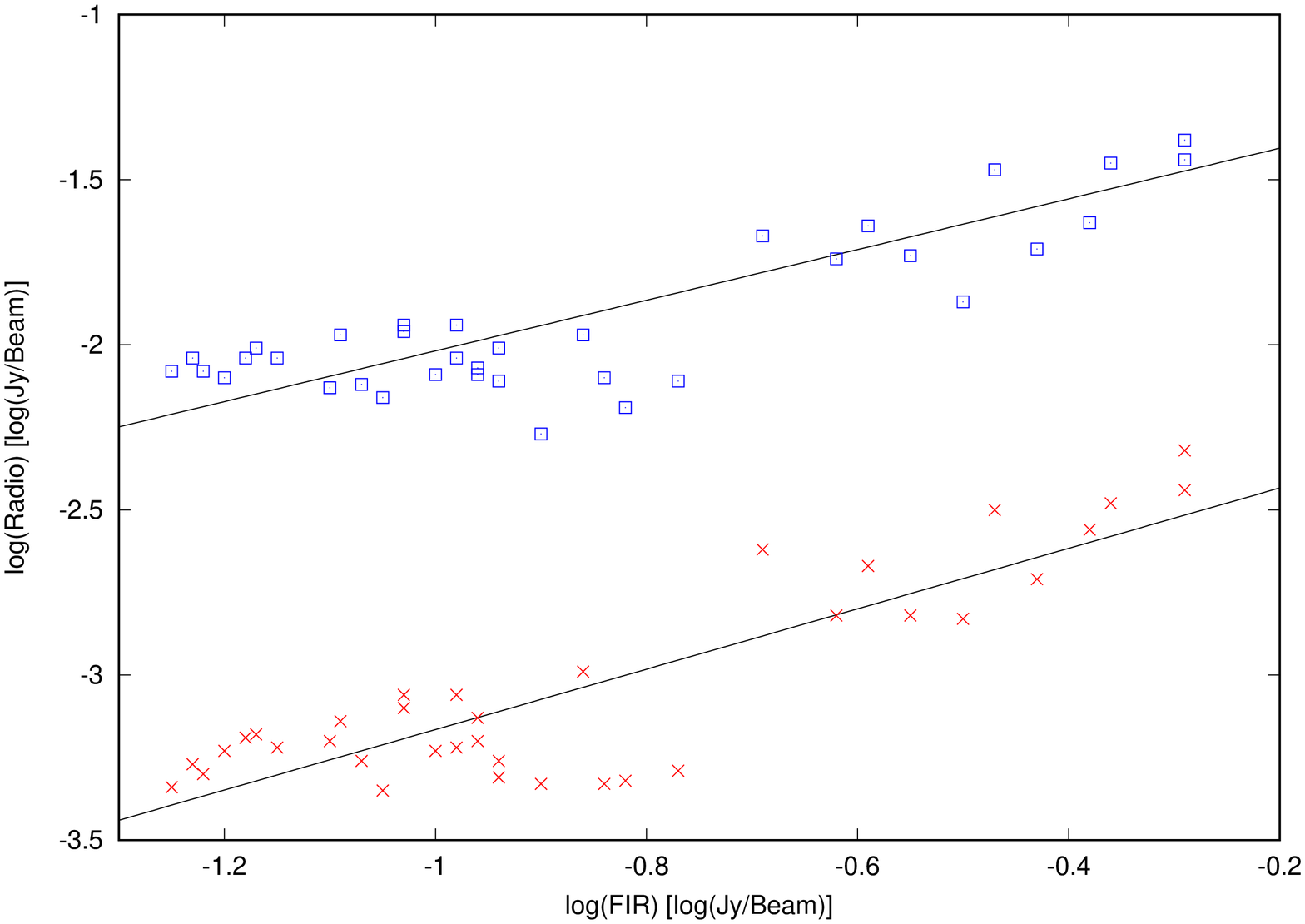}
\caption{70 $\mu$m FIR vs 0.33 and 1.4 GHz radio intensity (lower plot) in the interarm region of NGC 5194. 
To the 0.33 GHz data 0.5 dex has been added.}
\end{figure}

\begin{table*}
\caption{Fitted values of the radio-FIR correlation at 0.33 and 1.4 GHz of the sample galaxies.}

\centering
 \begin{tabular}{||c c c c c c c||} 
 \hline
  Name   & b (0.33 GHz)        & b (1.4 GHz)        & log(a) [0.33 GHz]   & log(a) [1.4 GHz] & Scatter (0.33 GHz) & Scatter (1.4 GHz)\\ 
         &                     &                    & [log(Jy/Beam)]     & [log(Jy/Beam)] & [log(Jy/Beam)] & [log(Jy/Beam)] \\
 \hline\hline
 NGC 2683 & 0.54 $\pm$ 0.06 & 0.91 $\pm$ 0.07 & -2.29 $\pm$ 0.04 & -2.47 $\pm$ 0.05 & 0.07 &       0.08\\
 NGC 3627 & 0.55 $\pm$ 0.03 & 0.85 $\pm$ 0.13 & -1.78 $\pm$ 0.02 & -2.5  $\pm$ 0.05 & 0.13 &       0.12\\
 NGC 4096 & 0.74 $\pm$ 0.05 & 0.90 $\pm$ 0.04 & -2.01 $\pm$ 0.04 & -2.35 $\pm$ 0.04 & 0.10 &       0.10\\
 NGC 4449 & 0.77 $\pm$ 0.05 & 0.65 $\pm$ 0.04 & -2.14 $\pm$ 0.03 & -2.35 $\pm$ 0.02 & 0.13 &       0.12 \\  
NGC 4490 & 0.68 $\pm$ 0.02 & 0.75 $\pm$ 0.02 & -1.62 $\pm$ 0.02 & -1.94 $\pm$ 0.01 & 0.12 &       0.11 \\
 NGC 4826 & 1.39 $\pm$ 0.1  & 1.47 $\pm$ 0.08 & -2.78 $\pm$ 0.05 & -2.93 $\pm$ 0.04 & 0.09 &       0.11\\
 NGC 5194 (arm) & 0.50 $\pm$ 0.05 & 0.65 $\pm$ 0.04  & -1.85 $\pm$ 0.02 & -2.23 $\pm$ 0.02 & 0.16 & 0.17\\ 
 NGC 5194 (interarm) & 0.73 $\pm$ 0.11 & 1.03 $\pm$ 0.05  & -1.74 $\pm$ 0.05 & -2.12 $\pm$ 0.04 & 0.15 & 0.19\\
 \hline
 \end{tabular}
\end{table*}

\section{Discussions}

\subsection{How cosmic ray electrons propagate in nearby galaxies}

As discussed in Sect. 4.3, CRes are expected to spread out from the sources of
emissions in a galaxy. Therefore, compared to FIR band, where the thermal
sources of emissions are mostly seen and which are believed to give rise to
CRes, radio image of a galaxy would appear to be smeared by a convolution
function (kernel). 
For the simple scenario of CRe diffusion,
ratio of FWHMs of these kernels is $\propto (\nu_1 / \nu_2)^{- 1 / 4}$. Therefore,
the ratio of FWHMs in this case is $\sim$1.4 between 0.33 and 1.4 GHz, and $\sim$2.1 between
0.33 and $\sim$6 GHz. For propagation through streaming instability, the ratio
is $\propto (\nu_1 / \nu_2)^{- 1 / 2}$, which is $\sim$2.1 between 0.33 and 1.4 GHz, and is 
$\sim$4.3 between 0.33 and $\sim$6 GHz. 
From  Table-3, we find that for NGC 4096 and 4490, the ratios of the FWHMs from P
and L/C band are consistent (given the errors) with the simple diffusion
scenario. For NGC 3627 and 5194, the ratio of convolution FWHM from the P-band
($\nu_1$) and C or/and L-band ($\nu_2$), is consistent with streaming
instability. This mechanism is also consistent with CRe
propagation scale found above from the 0.14 GHz thermal subtracted LOFAR image
(Sect. 4.3) for NGC 5194.  It is however at variance with the detailed modelling of
CRe propagation in this galaxy \citep{Mulcahy2016A&A}.
For NGC 2683, kernel FHWM in L and C band remains the same, but increases 
by a factor of 6 in P band. Reconciling these three results from a physically
intuitive scenario is not possible. Given the higher structural sensitivity of
the C band image as compared to the L band one, we only consider the ratio of
FWHMs from P and C band images from Table-3, which is consistent with CRe
propagation due to streaming instability. Further observations would be needed
to confirm the ratio of FWHMs of CRe propagation between L and P band for this galaxy.
For NGC 4449, the ratio of CRe propagation FWHMs from P and C band is $\sim$1,
suggestive of galactic outflow (diffusion could explain it if the measured ratio
(Table-3) is off by 2$\sigma$).
This could not be caused by lack of short {\it uv}-spacing or low sensitivity
in P band, but it needs to be verified by independent observations and
analysis.

It should be noted that diffusion is expected to happen in all galaxies at large length
scales ($\gtrsim$ tens of kpcs), but indications of the other mechanisms have
been suggested in literature.  Streaming instability is indicated in NGC 1097,
NGC 5055 and IC 342 \citep{Basu2013MNRAS,Beck2015A&A}. \citet{Heesen2018MNRAS}
present cases of 11 galaxies where advection (galactic outflows) plays the
dominant role. Diffusion is indicated in NGC 4736, NGC 5236, NGC 6946 and NGC
5194 \citep{Basu2013MNRAS,Mulcahy2016A&A}.  
We have made several assumptions on the properties of the convolution
kernel. While determining the FWHM of the convolution kernel, we
assumed that the exponent of radio-FIR correlation is unity, which is not true for
galaxies with significant escape of CRes. A more plausible solution would
involve model of a galaxy and numerical simulation involving different physical models of
CRe propagation and then finding a generic solution as a combination of all these
processes. While fitting the data for correlation between radio and
convolved FIR maps, we had to apply lower and in many cases upper cut-offs to
the data, which are somewhat subjective in nature. 
FIR images with higher resolution, better technique to analyse the images
and a larger sample size would be needed to get a better handle in unravelling the 
mechanisms of CRe propagation in nearby normal galaxies.
Such a process would be attempted in future involving a larger number of
galaxies from our original sample.

\subsection{Low frequency radio-FIR correlation}
CRes with lower energy propagate a longer distance in their lifetime than their
higher energy counterparts. In a typical magnetic field of 10 $\mu$G,
the peak emission at 1.4 GHz
would occur from electrons with energy $\sim$3 GeV, and has a typical lifetime
of $\sim 2.5 \times 10^7$ years. In this timescale, they could propagate to a
distance of $\sim1$ kpc. With the above magnetic fields for the 1.5 GeV
electrons, they would survive twice longer and could propagate to a distance of
$\sim2$~kpc during their lifetime. 
From our sample, the average exponent of the
correlation between 0.33 GHz radio and 70 $\mu$m FIR is $0.78\pm0.1$ (for NGC 5194,
we only considered the exponent from the arm regions). The galaxy NGC 4826 in the
sample has an anomalously high exponent of 1.39 at 0.33 GHz. 
As mentioned earlier in Sect.~4.3, a large part of the emission originates
from the core region of the galaxy that may have a different ratio of radio to
FIR emission, which would affect the exponent of
the radio-FIR correlation.
Also, there could be significant AGN contribution to the emission from the core.
Therefore, the exponent of radio-FIR correlation for this galaxy needs to be
used with caution. Its exponent of radio-FIR correlation well above unity is,
however, consistent with the non-calorimeter model \citep{Niklas1997A&A}. If
this galaxy is excluded from Table-4, the mean exponent
is $0.63\pm0.06$.  This exponent at 0.33 GHz is significantly lower than
0.78$\pm$0.04 obtained using 1.4 GHz data for the same galaxies.
Since CRes propagate significantly further than the linear resolution of our
observations at 0.33 GHz, this would cause the radio-FIR correlation to flatten
from 0.33 GHz radio maps as compared to 1.4 GHz, and can explain the above
result. However, we note that the above trend is not seen from 0.33
and 1.4 GHz maps of NGC 4449 (see Table-4).  As noted in Sect.~4.3,
1.4 GHz image of NGC 4449 made from VLA archival data
has poorer structural sensitivity than the 0.33 GHz image. Consequently, 'b' value measured
using the 1.4 GHz emission would arise from much smaller core regions of the 0.33 GHz
emission, and the two results cannot be compared. 
We also consider the exponent obtained from our 0.33 GHz observation with  \cite{Basu2012ApJ}. 
Considering that most of galactic emission in spirals originates from arm regions, 
we find from Table-3 of \cite{Basu2012ApJ} the 0.33 GHz radio to 70 $\mu$m FIR
average exponent is $0.57\pm 0.05$ for their sample of 4 galaxies.
 For
NGC 5194, spatially separated radio-FIR correlation for the arm and interarm
regions are shown in Fig.~36 and 37. Contrary to what was found by
\cite{Basu2012ApJ} for other large face-on galaxies, we do not observe any
significant flattening of the exponent for the interarm region for this galaxy at
0.33 and 1.4 GHz. At 0.33 GHz, the exponent for the interarm region is 0.73$\pm0.11$,
which is way steeper than the average exponent of $0.33\pm0.02$ found for the
interarm regions of the four nearby spirals \citep{Basu2012ApJ}. 

\section{Conclusions}
(i) We have observed a set of seven galaxies NGC 2683, NGC 3627, NGC 4096, NGC
4449, NGC 4490, NGC 4826 and NGC 5194 within 11 Mpc at 0.33 GHz using GMRT and
also analysed their archival higher radio frequency data at 1.4 or $\sim$6 GHz.
Their total intensity and non-thermal spectral index maps are presented here.
Radio maps for most of these galaxies were not available below 1 GHz with high
resolution ($\sim10^{''}$) and sub-mJy sensitivity.

(ii) The spectral indices of the galaxies are typically flat at the
centre and in compact star forming regions. It gradually steepen with
increasing galactocentric distances.

(iii) 
To determine the characteristic smearing scales due to propagation of CRes in
galaxies at different radio bands, we assumed a simple isotropic diffusion model,
and convolved the 70~$\mu$m FIR maps of the galaxies with Gaussian kernels with
different angular sizes till the exponent of their radio-FIR correlation reached
unity.
CRe propagation in NGC 4096 and 4490 could be explained by the simple diffusion scenario.
For NGC 2683, 3627 and 5194, CRe propagation through streaming instability could explain the
results. For NGC 4449, propagation through galactic outflow is suggested.

(iv) We have studied spatially resolved radio-FIR correlation between 0.33, 1.4 GHz
radio and 70 $\mu$m FIR emission.
The average exponent of the radio-FIR correlation for six of the galaxies is
found to be 0.63$\pm$0.06 using radio maps at 0.33 GHz, and is
quite close to the average exponent of $0.57\pm 0.05$ obtained for the arm regions of
4 nearby large spirals \citep{Basu2012ApJ}.  

\section*{Appendix}
\subsection*{CRe propagation from a compact source due to isotropic diffusion}

Outflow ($J$) at a distance `r' from a source can be written as
$J (r) = - D. \frac{\tmop{dn}}{\tmop{dr}}$, where $\frac{\tmop{dn}}{\tmop{dr}}$
is the net density gradient of CRes, and $D$ is the diffusion coefficient
considered constant.  
Also, $\frac{\delta n}{\delta t} = \frac{\delta J}{\delta r}$.
The solution of which is 
$$n (r, t) = \frac{N}{\sqrt{4 \pi \tmop{Dt}}}\exp \left( - \frac{r^2}{4 \tmop{Dt}} \right)$$
where, N is the rate of CRe generation from source.
After a long time (more than synchrotron lifetime) when steady state is reached,
observed density would be the integrated value of the above for all the CRe emission
over time from start ($t=0$) to the lifetime of synchrotron emission ($t=\tau_{\nu}$).

$$n_{\nu} (r) = \frac{2 N_{\nu}}{\sqrt{\pi D}} \int_0^{\tau_{\nu}}
\frac{1}{\sqrt{t}} \exp \left( \frac{- r^2}{4 \tmop{Dt}} \right) \tmop{dt}$$

where we have considered $N_{\nu}$ and $n_{\nu}$ are the rate of CRe production
and the density of CRes whose peak emission occurs at a frequency  $\nu$, and lifetime
of synchrotron emission is $\tau_{\nu}$.
The integrated value of the above equation is
$$n_{\nu} (r) = {\frac{2 N_{\nu}}{\sqrt{\pi D}}} \left( \sqrt{\tau_{\nu}}
\exp \left( - \frac{r^2}{4 D \tau_{\nu}} \right) - r \sqrt{\frac{\pi}{4 D}}
\tmop{erf} \left( \frac{r}{\sqrt{4 D \tau_{\nu}}} \right) \right)$$

Considering emission at 0.33 GHz, the typical astrophysical values of $r$ is
$\sim$kpc, $\tau_{\nu} \sim 10^8$ yr and $D \sim 10^{28} \tmop{cm}^2
\sec^{- 1}$. It can then be shown that typically the first term (before the
negative sign) dominates, which indicates a Gaussian profile of CRe density
distribution as a function of distance from the source.

\section*{Acknowledgements}
We thank Dipanjan Mitra for reading the manuscript and providing useful comments.
We also thank the anonymous referee for important comments which helped to
improve the quality of the paper and to avoid certain significant errors.  We
thank the staff of GMRT that allowed these observations to be made. GMRT is run
by National Centre for Radio Astrophysics of the Tata Institute of fundamental
research. We acknowledge support of the Department of Atomic Energy, Government
of India, under project no. 12-R\&D-TFR-5.02-0700.

\section*{Data Availability}

The raw interferometric data at 0.33 GHz used in this paper is publicly
available from the GMRT Online Archive at https://naps.ncra.tifr.res.in/goa
(Project: 29\_088).  Advanced data products used in the article will be shared
on reasonable request to the corresponding author.

\bibliographystyle{mnras}
\bibliography{nearby.galaxies.mod}
\label{lastpage}
\end{document}


\maketitle

\section*{Non-thermal spectral index images of NGC 2683, 3627, 4096 
and 4449 between 1.4 and 0.33 GHz and their thermal emission subtracted 1.4
GHz maps}

\vspace{1 cm}

\begin{figure}
   \centering
\includegraphics[width=\columnwidth]{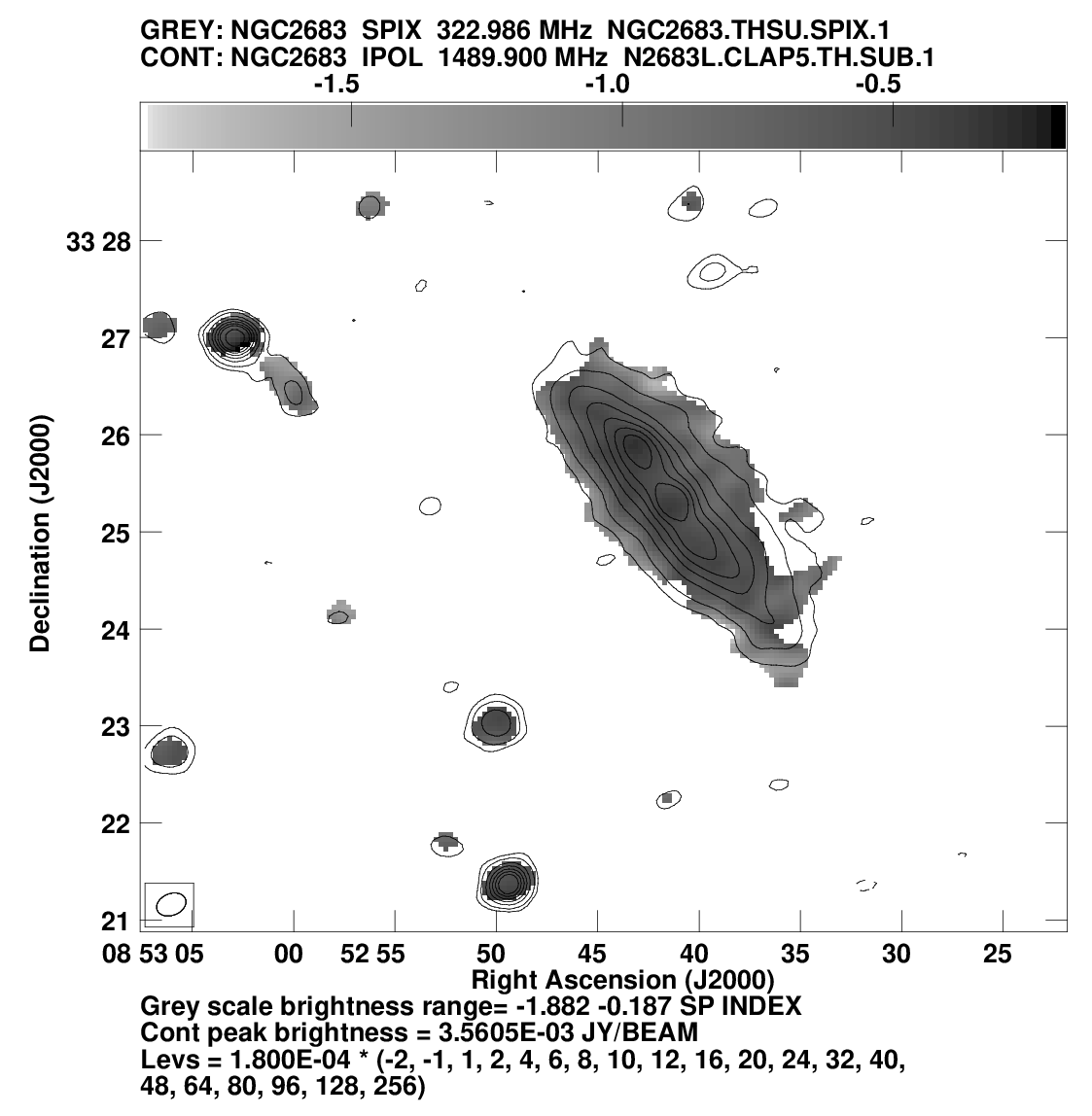}
\caption{Thermal emission subtracted 1.4 GHz map (contour) and non-thermal
spectral index of NGC 2683 (gray). Resolution is 19$^{''}\times 13^{''}$.}
\end{figure}.

\begin{figure}
   \centering
\includegraphics[width=\columnwidth]{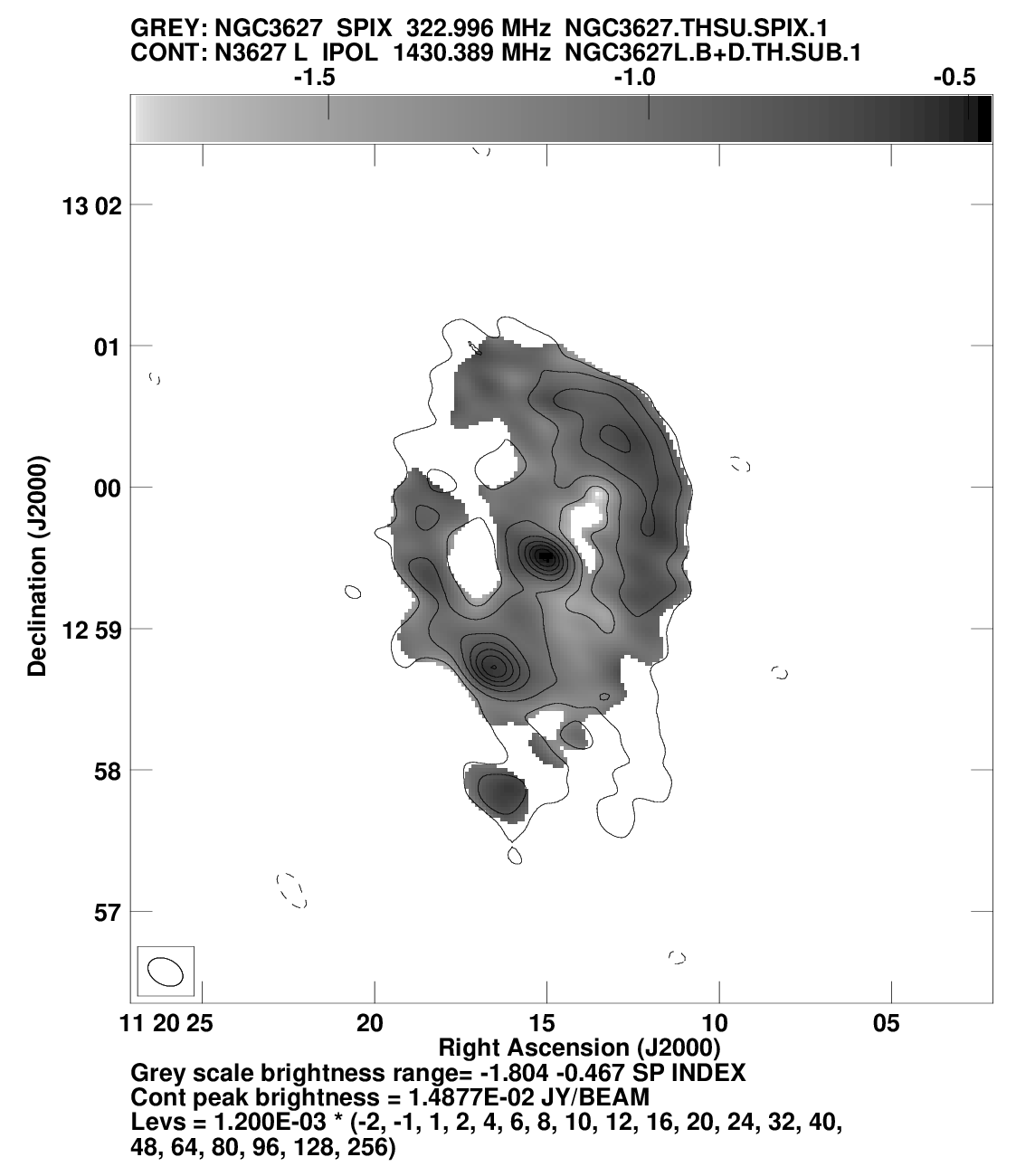}
\caption{Thermal emission subtracted 1.4 GHz map (contour) and non-thermal
spectral index of NGC 3627 (gray). Resolution is 16$^{''}\times 11^{''}$.}
\end{figure}

\begin{figure}
\includegraphics[width=\columnwidth]{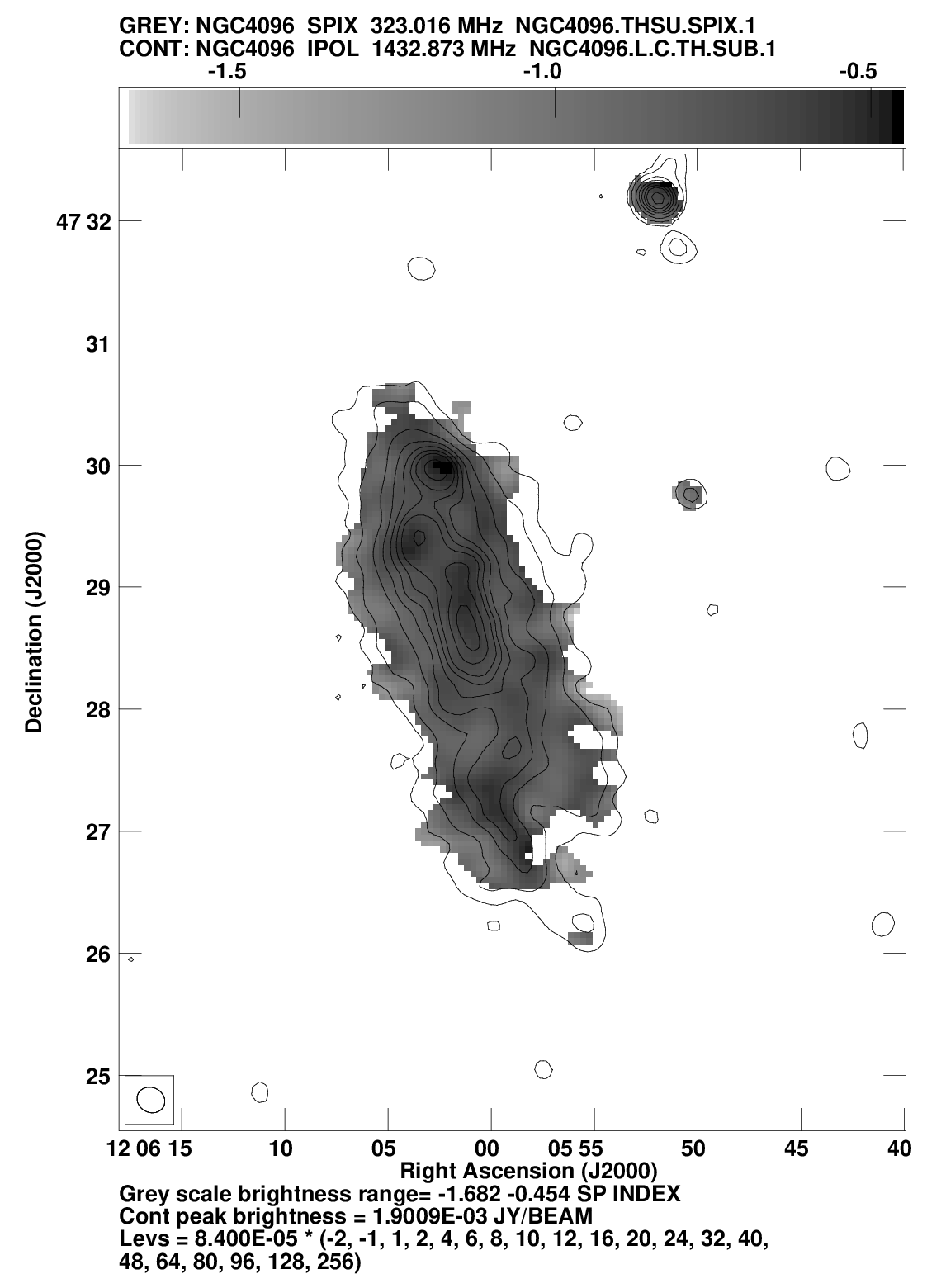}
\caption{Thermal emission subtracted 1.4 GHz map (contour) and non-thermal
spectral index of NGC 4096 (gray). Resolution is 14$^{''}\times 12^{''}$.}
\end{figure}

\begin{figure}
\includegraphics[width=\columnwidth]{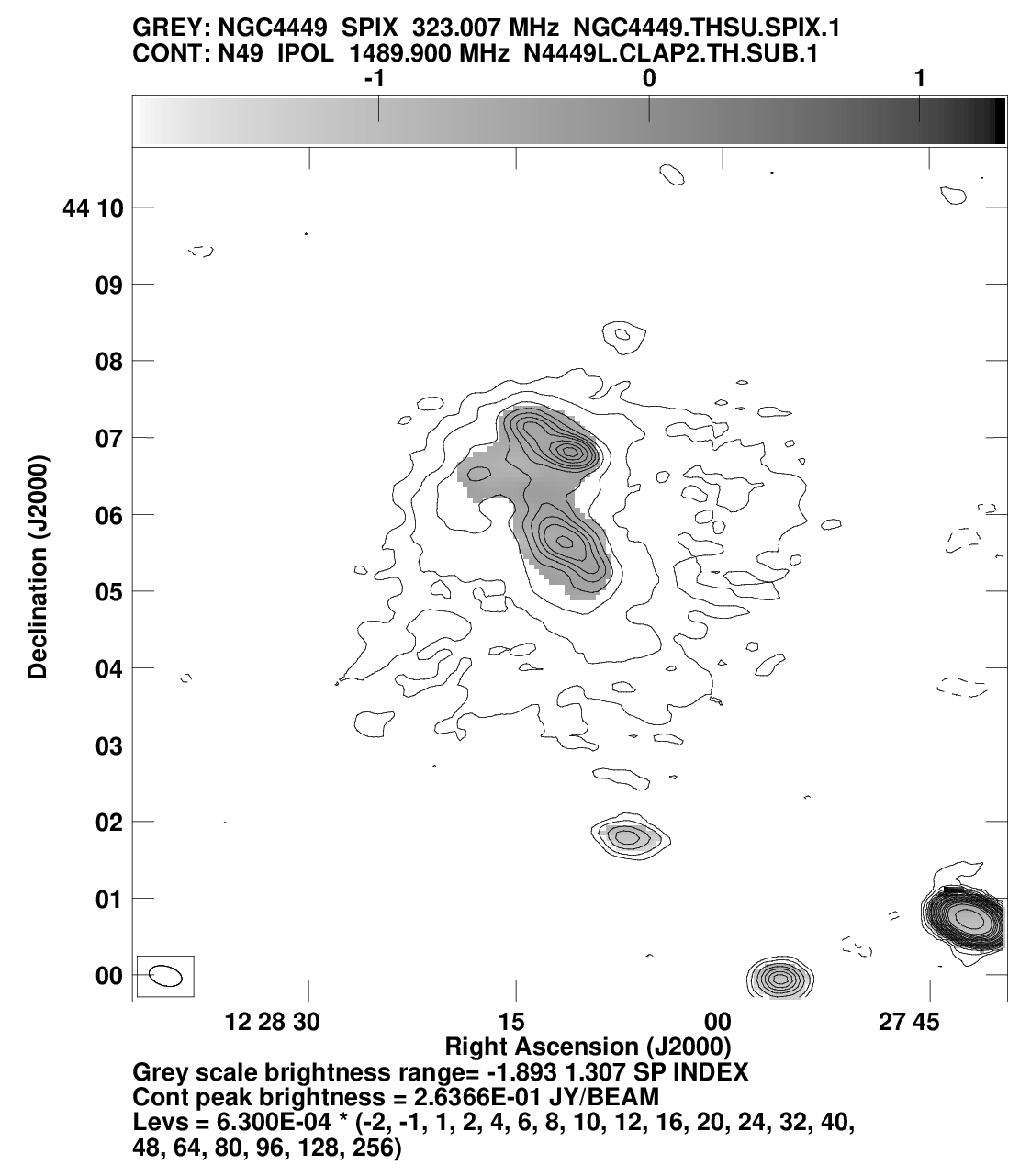}
\caption{Thermal emission subtracted 1.4 GHz map (contour) and non-thermal
spectral index of NGC 4449 (gray). Resolution is 26.1$^{''}\times 15^{''}$.}
\end{figure}
